\documentclass[journal]{IEEEtran}
\normalsize

\usepackage{bm}
\usepackage{multirow}

\usepackage{bbding}
\usepackage{amsmath,amssymb,amsfonts}
\usepackage{textcomp}
\usepackage{xcolor}
\usepackage{makecell}
\usepackage{url}

\usepackage{cite}

\ifCLASSINFOpdf
\else
\fi
\usepackage{graphicx}
\usepackage{booktabs}

\usepackage{amsthm}

\newtheoremstyle{nts1}
{}                
{}                
{\itshape}        
{\parindent}      
{\itshape}        
{:}               
{ }               
{\thmname{#1}\ \thmnumber{#2}\itshape\ \thmnote{(#3)}\itshape}                

\newtheoremstyle{nts2}
{}                
{}                
{\itshape}        
{\parindent}      
{\itshape}        
{:}               
{ }               
{}                

\theoremstyle{nts1}

\theoremstyle{nts2}



\makeatletter  
\newif\if@restonecol  
\makeatother

\usepackage[linesnumbered,ruled,vlined]{algorithm2e}
\usepackage{algpseudocode}  
\usepackage{amsmath}  

\SetAlgoCaptionSeparator{}

\usepackage{etoolbox}

\providetoggle{beginFlag}
\settoggle{beginFlag}{true}

\providetoggle{beginFlag}
\settoggle{beginFlag}{true}

\providetoggle{beginFlag}
\settoggle{beginFlag}{true}

\usepackage{array}
\usepackage{tabularx}

\usepackage{caption}
\usepackage{float}
\usepackage{subfigure}
\usepackage{setspace}

\usepackage{url}

\usepackage[hidelinks]{hyperref}

\begin{document}
%

\title{Scalable Full-Horizon SFC Deployment over Predictable SAGIN Topologies via a Configurable Column Generation Framework}

%
%
%


\author{Hang Li,
        Hao Huang,
        Yihu Xu,
        Zixuan Kong,
        and Yanling Jin
        \thanks{This research was funded by the Jilin Province Natural Science Foundation, grant number YDZJ202501ZYTS641.}
        \thanks{Hang Li and Yihu Xu are with the College of Engineering, Yanbian University, Yanji 133002, China (e-mail: lihang0402@gmail.com and lihang0402@ybu.edu.cn for Hang Li; xuyh@ybu.edu.cn for Yihu Xu). Hao Huang is with Chongqing University of Posts and Telecommunications, Chongqing 400065, China (e-mail: huanghao@cqupt.edu.cn). Zixuan Kong is with Beijing University of Posts and Telecommunications, Beijing 100876, China (e-mail: kong\_zx@bupt.edu.cn). Yanling Jin is with Fujitsu Limited, Japan (e-mail: kinkenrei83@gmail.com). (\em{Corresponding author: Yihu Xu}.)}
}

\maketitle

\begin{abstract}
Space--air--ground integrated networks (SAGINs) integrate heterogeneous communication and computing resources to support ubiquitous services. Service function chains (SFCs) enable complex services to be composed of ordered functions and executed collaboratively across these resources. Mobility continuously reshapes connectivity throughout service execution, yet much of the resulting topology evolution is predictable. Decisions based only on isolated topology snapshots may therefore overlook valuable future connectivity. We formulate an admission-first, cost-aware model on a time-expanded graph for SFC deployment. Our key observation is that temporal feasibility is request-specific, whereas global coupling arises mainly from shared-resource competition. We therefore develop a configurable column-generation framework in which complete request deployments form columns and a restricted master problem coordinates admission and resources. SAGINFE, a fast constructive heuristic built on per-flow embedding (PFE), supplies a standalone feasible solution and initial columns. Configurable pricing schedules combine early-terminated and stricter searches, supplemented by selective PFE candidates. Candidates undergo feasibility and exact reduced-cost checks. Request-wise pricing can run in parallel, followed by finite-pool integer selection. The default configuration improves the admission-first objective over initialization on all 120 main instances. Four pricing workers achieve up to a 2.69-fold median pricing-stage speedup while preserving admitted request counts in the tested cases.
\end{abstract}


%
\IEEEpeerreviewmaketitle

\section{Introduction}\label{sec:introduction}
%
%
%
%

\IEEEPARstart{S}{pace}--air--ground integrated networks (SAGINs) combine satellites, aerial platforms, and terrestrial infrastructure to extend network coverage and support services beyond the reach of ground networks. Besides communication connectivity, the computing resources distributed across these network segments allow service processing to be performed collaboratively \cite{liu2018sagin,zhou2019bidirectional}. Network function virtualization (NFV) facilitates this collaboration by implementing network functions as software that can be deployed on different computing nodes \cite{mijumbi2016nfv}. In an NFV-enabled network, a service can be represented by a service function chain (SFC), which specifies the functions that its data must traverse and their required processing order \cite{halpern2015sfc}. Deploying an SFC therefore requires jointly determining virtual network function (VNF) execution locations and the routing paths connecting them \cite{bari2016orchestrating}. This provides a flexible means of coordinating communication and computing resources for service provisioning in SAGINs \cite{wang2020sfc}.

Nevertheless, such flexibility is accompanied by additional deployment challenges. In SAGINs, the movement of satellites and aerial platforms changes network connectivity during service execution. A node suitable for processing a VNF in the current time slot may not provide an immediate connection to the node selected for the next VNF. Meanwhile, satellite orbits and planned aerial trajectories make much of this topology evolution predictable. Flow data can consequently be processed at an available node, stored there, and forwarded through a contact available in a later time slot \cite{araniti2015contact}. Decisions based only on isolated topology snapshots may overlook these opportunities. Exploiting predictable connectivity requires planning function execution, transmission, and waiting throughout each request's lifecycle, while ensuring that all required VNFs are executed in order and that the data reach the destination before the deadline.

Existing studies have advanced SFC deployment through joint resource orchestration, time-expanded modeling, and decomposition-based optimization \cite{wang2020sfc,jia2020virtual,jia2021vnf,yuan2024joint,jia2025service}. These developments provide a basis for using future connectivity in deployment planning. Their practical use also raises a computational question: how can a fast feasible deployment be followed by joint optimization whose search effort and execution can be configured for different problem sizes and computation requirements? Addressing this question requires connecting two aspects of full-horizon planning: the effort spent finding alternative deployments and the coordination of these alternatives under shared resource constraints.

The first challenge concerns the variable computational burden introduced by full-horizon planning. A time-expanded graph (TEG) replicates network nodes across time slots and represents transmission and waiting opportunities. Even with a fixed physical network, extending the planning horizon adds node copies and can enlarge the deployment search space. Network size, request count, chain lengths, and resource competition further affect the search burden. Decomposition reduces individual problem sizes, but repeated per-request searches can remain expensive. Different planning instances therefore call for different computational effort. This motivates fine-grained quality--time control through search strength and scheduling, supplementary candidates, and parallel execution within a common framework.

The second challenge is turning that search effort into a jointly feasible deployment. In a sequential construction procedure, an early placement or routing choice may consume resources needed by later requests. Improving admission can therefore require revising several deployment choices together. The initial deployments and newly generated alternatives need a common representation so that they can be jointly reconsidered without violating temporal feasibility or shared resource capacities. A feasible starting solution should remain available when additional search yields no better deployment.

In view of the above challenges, we study full-horizon SFC deployment over predictable SAGIN topologies. We consider offline planning with known topology evolution and request information, and formulate an integer linear program (ILP) that first maximizes the number of admitted requests and then minimizes normalized resource consumption. Each admitted request has a complete deployment from its source and arrival slot to its destination within the deadline, including ordered VNF execution, transmission, and waiting. Our design exploits the fact that temporal feasibility is request-specific, whereas the coupling among requests arises from shared node and link resources. Following the Dantzig--Wolfe decomposition principle \cite{vanderbeck2000dantzig}, we represent complete per-request deployments as columns. A restricted master problem (RMP) coordinates admission and resource allocation using these alternatives. This representation connects initial construction with subsequent joint optimization under the same objective and deployment constraints.

To obtain a feasible starting solution efficiently, we develop the SAGINFE heuristic. Its per-flow embedding (PFE) procedure searches successive VNF stages over the TEG while tracking remaining resources. SAGINFE provides a standalone deployment and initial per-request columns for subsequent joint optimization.

Each pricing subproblem seeks a complete deployment for one request under dual prices from the RMP's linear relaxation. Solvers and problem-specific heuristics can therefore serve as interchangeable candidate generators, subject to feasibility and exact reduced-cost checks before insertion. The same interface could also accommodate other methods that construct complete feasible deployments for individual requests.

The subsequent optimization provides fine-grained quality--time control through pricing schedules that arrange early-terminated and stricter solver-based searches, selective PFE candidate generation, and configurable request-wise parallelism. Finally, integer selection over the generated finite column pool yields compatible deployments. Comparing this result with the SAGINFE initialization preserves the better feasible solution under the admission-first objective.

Specifically, our contributions in this paper are summarized as follows:

\begin{enumerate}

\item We formulate full-horizon SFC deployment over predictable SAGIN topologies as an admission-first, cost-aware ILP and prove its NP-hardness. The model jointly describes ordered VNF execution, communication, and waiting from request arrival to delivery within a deadline. Complete per-request deployments provide a common column representation for fast construction and subsequent resource coordination.

\item We develop a configurable column-generation framework that connects SAGINFE initialization with subsequent joint optimization. Its per-request column interface accommodates solver-based and heuristic candidates. Pricing-stage schedules, selective candidate generation, and request-wise parallelism provide fine-grained control of computational effort while retaining the same deployment objective and feasibility requirements. Finite-pool integer selection and an initial-solution safeguard yield a feasible result no worse than the SAGINFE initialization under the stated objective.

\item We evaluate the framework across network scales, traffic loads, and planning horizons, including pricing-schedule comparisons and component ablations. The default configuration improves the admission-first objective over initialization on all 120 main instances. Parallel-pricing experiments demonstrate up to a 2.69-fold median pricing-stage speedup with four workers, without changing admitted request counts in the evaluated cases.

\end{enumerate}

The remainder of the paper is organized as follows. Section~\ref{sec:related-work} reviews related work, and Section~\ref{sec:system-model} presents the system model and problem formulation. Section~\ref{sec:framework} develops the proposed solution framework, which is evaluated through simulations in Section~\ref{sec:simulation-results}. Finally, Section~\ref{sec:conclusion} concludes the paper and outlines future work.

\section{Related Work}\label{sec:related-work}

In this section, we first review temporal SFC deployment in satellite networks and SAGINs, and then discuss decomposition and the control of computational effort.

\subsection{SFC Deployment in Time-Varying Satellite Networks and SAGINs}

Several studies address SFC deployment by coordinating heterogeneous communication and computing resources. Wang et al. \cite{wang2020sfc} jointly optimize admission, VNF placement, and virtual-link routing in reconfigurable SAGINs, considering service revenue, function sharing, and resource costs. Their formulation uses the current topology and residual resources within a reconfiguration period; unfinished services continue into subsequent periods. For dynamic Internet of vehicles (IoV) services, Li et al. \cite{li2022cost} incorporate VNF live migration, reinstantiation, and rescheduling, and develop Tabu search-based algorithms. Zhang et al. \cite{zhang2026adaptive} investigate online SFC embedding in SAGIN-MEC using graph reinforcement learning and feasible-action masking.

Time-expanded models represent transmission and storage opportunities across slots. Jia et al. \cite{jia2020virtual,jia2021vnf} construct a software-defined time-evolving graph for LEO satellite networks, with storage arcs connecting successive copies of a satellite. Cao et al. \cite{cao2023sfc} extend temporal resource representation to SAGINs and use task--node matching to continue unfinished VNF stages after storage. Jia et al. \cite{jia2025service} further study urgent-task scheduling through deep reinforcement learning, tracking unfinished transmission and processing across slots.

Other studies use function-stage flow models. Yang et al. \cite{yang2023space} jointly optimize VNF deployment and routing on a multi-functional TEG, with binary mission-completion decisions requiring the entire data amount to be delivered within its service window. Wang et al. \cite{wang2026sfc} preserve function order through a flexible function-time expanded graph and maximize delivered data volume for a given SFC. Both optimize divisible stage flows across links and slots. Thus, complete-mission acceptance in \cite{yang2023space} does not require unsplit stage routes. Our model instead assigns each admitted request one execution per required VNF and whole-request routes between consecutive stages.

Liu et al. \cite{liu2026vmr} aggregate topology and resource evolution into a virtual multi-resource storage time aggregated graph. Their offline and online orchestration algorithms concern persistent SFC services, including duration, migration, delay, and jitter. This reduces temporal representation complexity for service continuity. Our model instead concerns finite requests with arrival slots and delivery deadlines.

\subsection{Decomposition and Control of Computational Effort}

Decomposition-based methods address the computational complexity of SFC optimization by separating coupled decisions into smaller subproblems. Jia et al.~\cite{jia2020virtual,jia2021vnf} use Dantzig--Wolfe decomposition and column generation for VNF provisioning in time-varying LEO satellite networks, with task-specific service paths as columns. Their formulation couples VNF placement with routing but does not explicitly enforce the execution order of successive VNFs.

Yuan et al. \cite{yuan2024joint} develop Benders branch-and-cut for latency minimization and a separate time expansion-based decoupled greedy algorithm. The latter constructs candidate paths before placing VNFs, with the candidate-path count controlling its quality--time tradeoff. Yang et al. \cite{yang2023space} instead separate groups of binary decisions and combine penalty-based optimization with linear programming (LP) relaxation and iterative rounding.

Building on these developments, we connect fast construction with configurable subsequent joint optimization for full-horizon, admission-first planning. Complete-request columns retain each request's ordered execution, transmission, and waiting requirements, while the RMP jointly selects admission or rejection alternatives under shared-resource constraints. SAGINFE supplies both an immediately usable deployment and initial columns; alternative feasible initializers and pricing generators can use the same interfaces. Pricing-stage schedules, selective heuristic candidates, and request-wise parallel execution configure subsequent computation without changing the model. Our contribution lies in the admission-first full-horizon formulation and its solution framework, which combines stage-wise feasible construction with configurable pricing and finite-pool integer selection. The evaluation examines how these components affect admission, resource cost, and runtime.


\section{System Model}\label{sec:system-model}

In this section, we introduce the network model and flow requests, and formulate the full-horizon SFC deployment problem over predictable SAGIN topologies.

\subsection{Network Model}\label{subsec:sm-network}

\subsubsection{Time Horizon and Nodes}

\begin{figure}[t]
\captionsetup{font={small}}
\centering
\includegraphics[width=1\columnwidth,height=1\textheight,keepaspectratio]{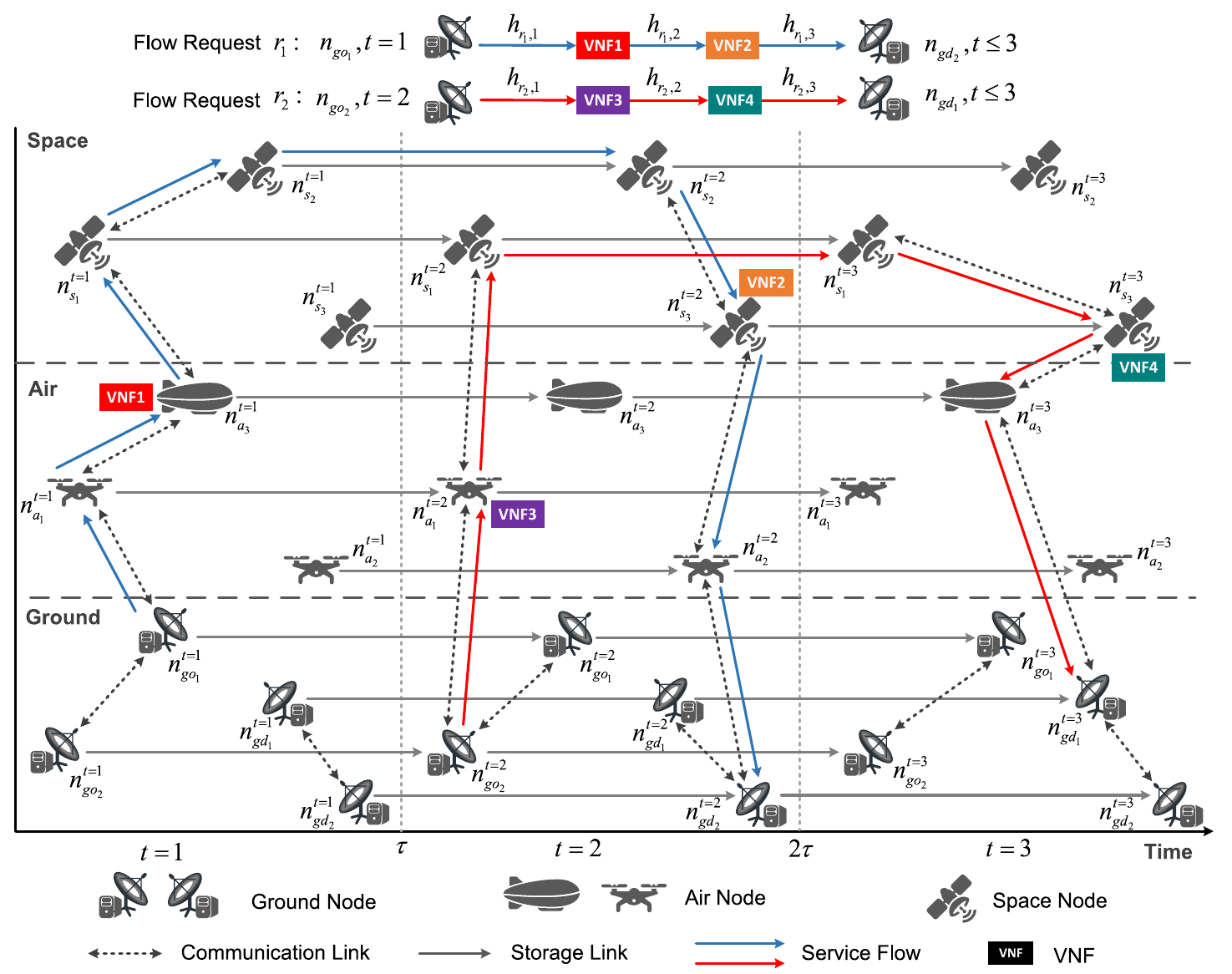}
\caption{An illustrative example of full-horizon, order-preserving SFC deployment over predictable time-varying SAGIN topologies.}\label{fig:sagin_example}
\end{figure}

We consider a SAGIN consisting of ground, air, and space nodes as shown in Fig.~\ref{fig:sagin_example}, whose physical node sets are denoted by $N_g$, $N_a$, and $N_s$, respectively. Considering the time-varying network connectivity, we focus on a time horizon $T=\{1,2,\ldots,|T|\}$, which includes $|T|$ time slots, and each time slot $t\in T$ has a duration of $\tau$. The network topology is regarded as quasi-static within each time slot, while its connectivity may change between time slots. Following the slot-based temporal representations in \cite{jia2021vnf,cao2023sfc,yuan2024joint}, we assume that $\tau$ is sufficient for the delays of the operations scheduled within a time slot. 

Then we build the corresponding time-expanded graph (TEG) of the SAGIN. Let $G=(N^T,E^T)$ denote the TEG, where $N^T$ and $E^T$ are the sets of time-expanded nodes and links, respectively. The node set is given by $N^T=N_g^T\cup N_a^T\cup N_s^T$, where $N_g^T$, $N_a^T$, and $N_s^T$ represent the time-expanded ground, air, and space node sets, respectively. Specifically, for each $t\in T$, every node $n\in N_g\cup N_a\cup N_s$ is represented by a corresponding time-indexed copy $n^t\in N^T$ in the time-expanded graph. Each time-expanded node $n^t\in N^T$ has a computing resource capacity $c(n^t)$, which represents the computing resource available at the corresponding physical node in time slot $t$.

\subsubsection{Communication and Storage Links}

The link set consists of communication links and storage links, i.e., $E^T=E_{\mathrm{com}}^T\cup E_{\mathrm{sto}}^T$. Communication links are bidirectional and connect node copies within the same time slot. Their directions are explicitly represented in the formulation, and the communication-link set is defined as $E_{\mathrm{com}}^T=\{e_{n^t m^t}^{t}\mid n,m\in N_g\cup N_a\cup N_s,\ n\ne m,\ t\in T\}$.

Storage links are unidirectional and connect the copies of the same physical node in two consecutive time slots. The storage-link set is defined as $E_{\mathrm{sto}}^T=\bigl\{e_{n^t n^{t+1}}^{t}\mid n\in N_g\cup N_a\cup N_s,\ t\in T\setminus\{|T|\}\bigr\}$.

Note that a communication link may not be available in every time slot. With node trajectories known over the planning horizon, link availability is known before planning. We use a binary parameter $\mu(e^t)$ to indicate whether link $e^t\in E^T$ exists in time slot $t$: $\mu(e^t)=1$ if the link exists, and $\mu(e^t)=0$ otherwise. Obviously, $\mu(e^t)=1$ for every storage link $e^t\in E_{\mathrm{sto}}^T$. For the later formulation, we define the neighbor-node set $\Phi_{n^t}$ of each $n^t\in N^T$ as $\Phi_{n^t}=\bigl\{m^t\mid \mu(e_{m^t n^t}^{t})=1\lor\mu(e_{n^t m^t}^{t})=1,\ m^t\in N^T\bigr\},\quad n^t\in N^T,\;t\in T$.

We use $c(e^t)$ to denote the bandwidth capacity of link $e^t\in E^T$, expressed as an effective data rate, with the uplink and downlink directions considered separately\footnote{Each bidirectional communication connection is represented by two directed links with possibly unequal capacities \cite{katz2003traffic}. Across all network segments, these capacities are assumed to result from a predetermined, jointly feasible resource allocation that accounts for duplexing, interference, and resource sharing. The effective rates incorporate allocated frequency resources or transmission-time shares \cite{fcc2025starlink,jeong2018mobile,3gpp2024rrc} and serve as known inputs to SFC orchestration.}. Following \cite{jia2021vnf}, storage capacity is assumed to be sufficient\footnote{As in \cite{jia2021vnf}, we assume sufficient storage capacity since storage resources are relatively inexpensive. This assumption can be readily relaxed by imposing finite capacities on storage links, with minor modifications to the algorithms proposed below.}, and hence $c(e^t)=\infty$ for $e^t\in E_{\mathrm{sto}}^T$. Accordingly, the maximum amount of data that link $e^t\in E^T$ can carry in time slot $t$ is $d(e^t)=\mu(e^t)\cdot c(e^t)\cdot\tau$. In particular, an unavailable communication link has $d(e^t)=0$, whereas a storage link has $d(e^t)=\infty$.

\subsection{Flow Requests}\label{subsec:sm-requests}

Let $R$ denote the set of flow requests. We represent each flow request $r\in R$ as a five-tuple $r=\left\{n_{r,o}^{t_{r,o}},\;n_{r,d}^{t_{r,d}^{\max}},\;V_r,\;H_r,\;f_r\right\}$. In particular, $n_{r,o}^{t_{r,o}}$ indicates that the data of $r$ become available for transmission at node $n_{r,o}\in N_g\cup N_a\cup N_s$ in time slot $t_{r,o}$. The destination term $n_{r,d}^{t_{r,d}^{\max}}$ specifies the receiving node $n_{r,d}\in N_g\cup N_a\cup N_s$ and its latest allowable receiving time slot $t_{r,d}^{\max}$. The actual receiving time slot is denoted by $t_{r,d}$, which satisfies $t_{r,o}\le t_{r,d}\le t_{r,d}^{\max}$.

The ordered VNFs required by $r$ constitute its SFC, denoted by $V_r=\{v_{r,1},v_{r,2},\ldots,v_{r,|V_r|}\}$, where $v_{r,k}$ is the $k$-th VNF in the chain. For a better description in the later formulation, we regard the source and actual destination as dummy VNFs, and set $n_{r,o}^{t_{r,o}}=v_{r,0}$ and $n_{r,d}^{t_{r,d}}=v_{r,|V_r|+1}$. Then the SFC is re-denoted as $\hat V_r=\{v_{r,0},v_{r,1},v_{r,2},\ldots,v_{r,|V_r|},v_{r,|V_r|+1}\}$. The flow length includes the two endpoint dummy VNFs and is therefore $|\hat V_r|=|V_r|+2$, whereas $|V_r|$ counts only the processing VNFs. For each $v_{r,k}\in\hat V_r$, where $k=0,1,\ldots,|V_r|+1$, we use $\delta_{v_{r,k}}$ to represent the node computing resource consumed by $v_{r,k}$ to process one unit of flow data. The two dummy VNFs do not consume computing resources, i.e., $\delta_{v_{r,0}}=\delta_{v_{r,|V_r|+1}}=0$.

The virtual-link set is denoted by $H_r=\{h_{r,1},h_{r,2},\ldots,h_{r,|V_r|+1}\}$, where each $h_{r,l}\in H_r$, with $l=1,\ldots,|V_r|+1$, represents the logical link connecting two adjacent VNFs $v_{r,l-1}$ and $v_{r,l}$ in $\hat V_r$. Finally, $f_r$ is the amount of data that must be processed and delivered for flow request~$r$.

Fig.~\ref{fig:sagin_example} presents an illustrative TEG over three time slots and the deployments of two flow requests. Request $r_1$ arrives in time slot $t=1$ and requires VNF1 followed by VNF2, while request $r_2$ arrives in time slot $t=2$ and requires VNF3 followed by VNF4. Both requests have a receiving deadline of $t=3$. Their colored trajectories show how communication, VNF execution, and storage are combined to complete the required chains in order. The storage segments also illustrate that a flow can remain at a node and continue its deployment using connectivity available in a later time slot.

\subsection{Decision Variables}\label{subsec:sm-variables}

We first introduce the flow admission decision variable

\begin{equation}
\label{eq:sm-admission-domain}
x_r\in\{0,1\},\quad\forall r\in R,
\end{equation}

which takes the value of $1$ if request $r$ is successfully deployed and $0$ otherwise. We then define the VNF deployment decision variable

\begin{equation}
\label{eq:sm-vnf-domain}
\begin{gathered}
y_{n^t}^{v_{r,k}}\in\{0,1\},\\
\forall r\in R,\;v_{r,k}\in\hat
V_r,\;n^t\in N^T,\;t\in T,
\end{gathered}
\end{equation}

where $k=0,1,\ldots,|V_r|+1$. It takes the value of $1$ if VNF $v_{r,k}$ is deployed at time-expanded node $n^t$, and $0$ otherwise.

For virtual-link deployment, we define the following binary variables for communication links and storage links, respectively:

\begin{equation}
\label{eq:sm-communication-domain}
\begin{gathered}
z_{e_{m^t n^t}^{t}}^{h_{r,l}}\in\{0,1\},\\
\forall r\in R,\;h_{r,l}\in H_r,\;e_{m^t
n^t}^{t}\in E_{\mathrm{com}}^T,\;t\in
T,
\end{gathered}
\end{equation}

\begin{equation}
\label{eq:sm-storage-domain}
\begin{gathered}
z_{e_{n^t
n^{t+1}}^{t}}^{h_{r,l}}\in\{0,1\},\\
\forall r\in R,\;h_{r,l}\in H_r,\;e_{n^t
n^{t+1}}^{t}\in E_{\mathrm{sto}}^T,\;t\in
T/\{|T|\},
\end{gathered}
\end{equation}

where $l=1,\ldots,|V_r|+1$. Each variable takes the value of $1$ if the corresponding link is used to deploy $h_{r,l}$, and $0$ otherwise. For brevity, $z_{e^t}^{h_{r,l}}$ refers to the corresponding variable for either link type.

A virtual link can only use a link available in the corresponding time slot. Therefore, the link deployment variables must satisfy

\begin{equation}
\label{eq:sm-link-availability}
z_{e^t}^{h_{r,l}}\le\mu(e^t),\quad
\forall r\in R,\;h_{r,l}\in H_r,\;e^t\in
E^T.
\end{equation}

\subsection{Deployment Constraints}\label{subsec:sm-constraints}

\subsubsection{Source and Destination Constraints}

Each admitted flow must start at its specified source in its arrival time slot and reach its specified destination no later than its deadline. Accordingly, we have

\begin{equation}
\label{eq:sm-source}
y_{n_{r,o}^{t_{r,o}}}^{v_{r,0}}=x_r,\quad\forall
r\in R,
\end{equation}

\begin{equation}
\label{eq:sm-destination}
\sum_{t'=t_{r,o}}^{t_{r,d}^{\max}}
y_{n_{r,d}^{t'}}^{v_{r,|V_r|+1}}=x_r,\quad\forall
r\in R.
\end{equation}

\subsubsection{VNF Processing Constraints}

Each admitted flow must be processed by every required VNF in its SFC exactly once. We therefore impose

\begin{equation}
\label{eq:sm-vnf-once}
\sum_{n^t\in
N^T}y_{n^t}^{v_{r,k}}=x_r,\quad
\forall r\in R,\;v_{r,k}\in\hat V_r.
\end{equation}

\subsubsection{Non-Splitting Constraints}

For each virtual link of a request, the flow cannot split over multiple outgoing links. Let $\mathbf{1}\{A\}$ denote the indicator function, which equals $1$ if condition $A$ holds and $0$ otherwise. For notational convenience, storage terms referring to nonexistent boundary links are defined as zero, without introducing additional decision variables. Since the last time slot has no outgoing storage link, we impose

\begin{equation}
\label{eq:sm-outgoing}
\begin{gathered}
\sum_{m^t\in\Phi_{n^t}}z_{e_{n^t m^t}^{t}}^{h_{r,l}}
+\mathbf{1}\{t<|T|\}z_{e_{n^t n^{t+1}}^{t}}^{h_{r,l}}\le 1,\\
\forall r\in R,\;h_{r,l}\in H_r,\;n^t\in N^T,\;t\in T.
\end{gathered}
\end{equation}

Similarly, at most one incoming link can be selected for each virtual link at a time-expanded node. Since no incoming storage link exists in the first time slot, we have

\begin{equation}
\label{eq:sm-incoming}
\begin{gathered}
\sum_{m^t\in\Phi_{n^t}}z_{e_{m^t n^t}^{t}}^{h_{r,l}}
+\mathbf{1}\{t>1\}z_{e_{n^{t-1}n^t}^{t-1}}^{h_{r,l}}\le 1,\\
\forall r\in R,\;h_{r,l}\in H_r,\;n^t\in N^T,\;t\in T.
\end{gathered}
\end{equation}

\subsubsection{Flow Conservation Constraints}

For each virtual link $h_{r,l}$, the routed data must connect the node executing $v_{r,l-1}$ to the node executing $v_{r,l}$. Using the indicator function to account for the time-slot boundaries, we have

\begin{equation}
\label{eq:sm-conservation}
\begin{gathered}
\begin{aligned}
&\sum_{m^t\in\Phi_{n^t}}
\left(z_{e_{m^t n^t}^{t}}^{h_{r,l}}-z_{e_{n^t m^t}^{t}}^{h_{r,l}}\right)\\
&\quad+\mathbf{1}\{t>1\}z_{e_{n^{t-1}n^t}^{t-1}}^{h_{r,l}}
-\mathbf{1}\{t<|T|\}z_{e_{n^t n^{t+1}}^{t}}^{h_{r,l}}\\
&\quad=y_{n^t}^{v_{r,l}}-y_{n^t}^{v_{r,l-1}},
\end{aligned}\\
\forall r\in R,\;h_{r,l}\in H_r,\;n^t\in N^T,\;t\in T.
\end{gathered}
\end{equation}

Together with the VNF processing constraints, this constraint connects consecutive VNFs in the required order.

\subsubsection{Resource Capacity Constraints}

The total computing resource consumed by all deployed VNFs at each time-expanded node must not exceed its computing capacity. Therefore, we have

\begin{equation}
\label{eq:sm-node-capacity}
\begin{gathered}
\sum_{r\in R}\sum_{v_{r,k}\in\hat V_r}
f_r\cdot\delta_{v_{r,k}}\cdot
y_{n^t}^{v_{r,k}}\le c(n^t),\\
\forall n^t\in N^T,\;t\in T.
\end{gathered}
\end{equation}

Similarly, the total data carried by a communication link in a time slot must not exceed its maximum transmissible data amount:

\begin{equation}
\label{eq:sm-link-capacity}
\begin{gathered}
\sum_{r\in R}\sum_{h_{r,l}\in H_r}
f_r\cdot z_{e^t}^{h_{r,l}}\le d(e^t),\\
\forall e^t\in
E_{\mathrm{com}}^T,\;t\in T.
\end{gathered}
\end{equation}

\subsection{Optimization Objective}\label{subsec:sm-objective}

We consider both node computing and link communication resource consumption when evaluating the deployment cost. To combine these two types of resources, we normalize their consumption by the corresponding total resource capacities over the time horizon. In the following expressions, a node $n^t$ or link $e^t$ in a summation indexed by $t$ refers to its copy in that time slot.

The total node computing capacity and total communication-link data capacity used for normalization are assumed to be positive and finite.

The normalized node resource consumption is calculated as

\begin{equation}
\label{eq:sm-node-cost}
C_{\mathrm{node}}=
\frac{\displaystyle
\sum_{r\in R}\sum_{t\in
T}\sum_{v_{r,k}\in\hat
V_r}
\sum_{n^t\in N^T}
f_r\cdot\delta_{v_{r,k}}\cdot
y_{n^t}^{v_{r,k}}}
{\displaystyle\sum_{t\in T}\sum_{n^t\in
N^T}c(n^t)}.
\end{equation}

By the node capacity constraints, we have $0\le C_{\mathrm{node}}\le 1$.

Similarly, the normalized link resource consumption is calculated as

\begin{equation}
\label{eq:sm-link-cost}
C_{\mathrm{link}}=
\frac{\displaystyle
\sum_{r\in R}\sum_{t\in T}\sum_{h_{r,l}\in
H_r}
\sum_{e^t\in E^T/E_{\mathrm{sto}}^T}
f_r\cdot z_{e^t}^{h_{r,l}}}
{\displaystyle
\sum_{t\in T}\sum_{e^t\in
E^T/E_{\mathrm{sto}}^T}d(e^t)}.
\end{equation}

Here, storage links are excluded from both the numerator and the denominator. By the link capacity constraints, we have $0\le C_{\mathrm{link}}\le 1$.

Now we formulate the SFC deployment problem as an integer linear program (ILP) with the following objective:

\begin{equation}
\label{eq:sm-objective}
\max\quad M\cdot\sum_{r\in R}x_r
-\psi\cdot C_{\mathrm{node}}
-(1-\psi)\cdot C_{\mathrm{link}},
\end{equation}

subject to the decision-variable domains \eqref{eq:sm-admission-domain}--\eqref{eq:sm-storage-domain}, link availability constraint~\eqref{eq:sm-link-availability}, and deployment constraints~\eqref{eq:sm-source}--\eqref{eq:sm-link-capacity}. In objective function~\eqref{eq:sm-objective}, the first term is the weighted number of admitted flow requests, where $M>1$ is chosen to give priority to flow admission. The second and third terms are the weighted normalized node and link resource consumption, respectively, where $0<\psi<1$ is the weight parameter. Since $0\le\psi C_{\mathrm{node}}+(1-\psi)C_{\mathrm{link}}\le 1$, admitting one additional request improves the objective by at least $M-1>0$, even under the worst-case increase in resource cost. Therefore, the objective first maximizes flow admission and then favors lower resource consumption among solutions admitting the same number of requests.

\medskip

\noindent\textbf{Proposition 1.} \emph{The full-horizon SFC deployment problem formulated above is NP-hard.}

\noindent\emph{Proof:} We prove NP-hardness by a reduction from 3-PARTITION, which is strongly NP-complete \cite{garey1978strong}. Given $3m$ positive integers $a_1,\ldots,a_{3m}$ and a positive integer $B$ satisfying $B/4<a_i<B/2$ and $\sum_{i=1}^{3m}a_i=mB$, this problem asks whether the integers can be partitioned into $m$ triples, each summing to $B$.

We construct a single-slot instance with $T=\{1\}$, $\tau=1$, and $N_a=N_s=\emptyset$, and fix $M=2$ and $\psi=1/2$. The ground network consists of $m$ processing nodes, each with computing capacity $B$, a common source, and a distinct common destination; both endpoints have zero computing capacity. The only communication links connect each endpoint to every processing node in both directions; each directed link has $\mu(e^1)=1$ and $c(e^1)=d(e^1)=6m$. For each $a_i$, we create a request $r_i$ with the common endpoints, $t_{r_i,o}=t_{r_i,d}^{\max}=1$, one required VNF, $f_{r_i}=1$, and $\delta_{v_{r_i,1}}=a_i$. By \eqref{eq:sm-vnf-once} and \eqref{eq:sm-node-capacity}, every admitted VNF must be assigned to exactly one processing node. Any assignment satisfying the node capacities admits a feasible routing: the two virtual links use the direct source-to-processing and processing-to-destination links, whose capacities suffice for all requests.

Consequently, all $3m$ requests can be admitted if and only if the integers can be assigned to the $m$ processing nodes without exceeding capacity $B$ at any node. Since the total demand is $mB$ and $B/4<a_i<B/2$, every node must then receive exactly three demands summing to $B$. Thus, all requests are admissible if and only if the 3-PARTITION instance is feasible. The construction has polynomial size. Since $M>1$ guarantees admission priority, a polynomial-time algorithm for the formulated optimization problem would determine whether all $3m$ requests can be admitted and hence solve 3-PARTITION. Therefore, the formulated problem is NP-hard. \hfill$\square$

\begin{figure*}[t]
\captionsetup{font={small}}
\centering
\includegraphics[width=0.9\textwidth,height=1\textheight,keepaspectratio]{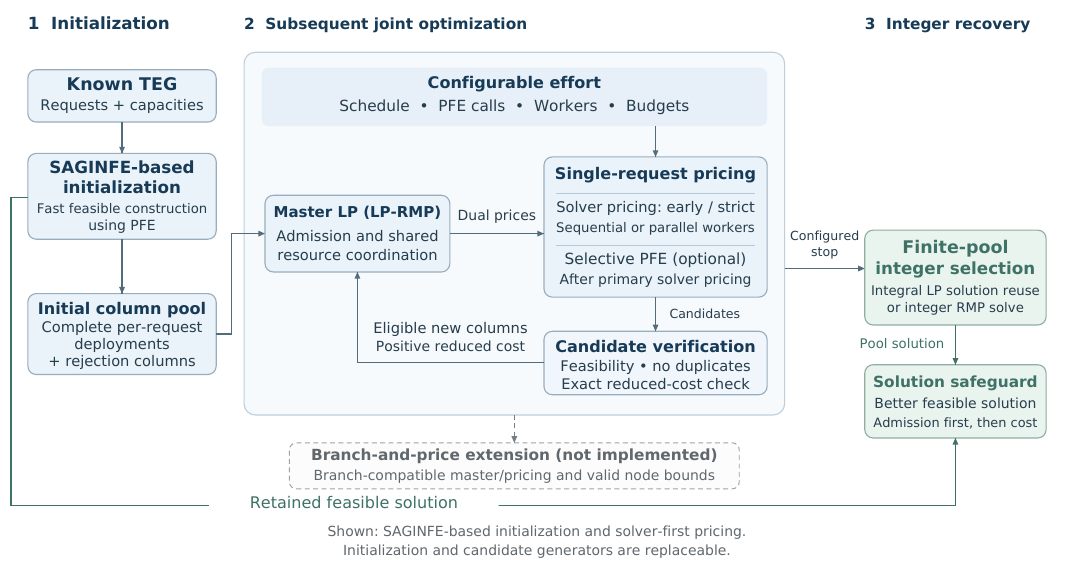}
\caption{Overview of the configurable column generation framework. The illustrated workflow uses SAGINFE initialization and solver-first pricing with optional PFE candidates; initialization and candidate-generation methods can be replaced. Solid arrows show the implemented workflow, whereas the dashed branch denotes a possible branch-and-price extension not implemented in this work.}
\label{fig:framework}
\end{figure*}

The node and link capacity constraints couple requests through shared resources, while the remaining deployment constraints describe each request's temporal feasibility. This structure motivates the request-wise column representation used in the next section.

\section{Configurable Column Generation Framework}\label{sec:framework}

In this section, we present a configurable column-generation framework that connects fast feasible construction with subsequent joint optimization.

\subsection{Framework Overview}\label{subsec:framework-overview}

The formulation in Section~\ref{sec:system-model} has a block-angular structure. This structure motivates a Dantzig--Wolfe decomposition \cite{dantzig1960decomposition,vanderbeck2000dantzig}: each column represents a complete deployment of one request, and the master problem coordinates the selection of these columns.

Fig.~\ref{fig:framework} organizes the workflow into initialization, subsequent joint optimization, and integer recovery. In the illustrated workflow, SAGINFE first constructs a feasible deployment and supplies its per-request deployments as initial columns. Starting from these columns and a rejection column for each request, the linear programming (LP) relaxation of the restricted master problem (RMP) produces dual prices for request-wise pricing. When the configured stopping rule is met, integer selection is performed over the generated finite column pool, and the better feasible solution is retained under \eqref{eq:sm-objective}.

The complete-deployment representation allows the framework to accommodate alternative initialization and pricing methods through modular interfaces. Initial columns must satisfy request-level feasibility, while a starting solution retained for comparison must additionally respect shared-resource capacities.

The decomposition can also serve as the basis of a branch-and-price extension \cite{barnhart1998branch,vanderbeck2000dantzig}, indicated by the dashed branch in Fig.~\ref{fig:framework}. This extension would require branch-compatible master and pricing problems and is not implemented here.

\subsection{Request-Based Dantzig--Wolfe Decomposition}\label{subsec:dw-decomposition}

\subsubsection{Complete Request Deployments}

For each request $r\in R$, let $\mathbf{y}_r$ and $\mathbf{z}_r$ collect its VNF execution and virtual-link deployment variables, respectively. An admitted deployment specifies $(x_r,\mathbf{y}_r,\mathbf{z}_r)$ with $x_r=1$ and satisfies the request-specific constraints \eqref{eq:sm-admission-domain}--\eqref{eq:sm-conservation}. We additionally retain the following single-request resource bounds:

\begin{subequations}\label{eq:dw-local-capacity}
\begin{align}
\sum_{v_{r,k}\in\hat V_r} f_r\delta_{v_{r,k}}y_{n^t}^{v_{r,k}} &\le c(n^t), &&\forall n^t\in N^T,\label{eq:dw-local-node}\\
\sum_{h_{r,l}\in H_r} f_r z_{e^t}^{h_{r,l}} &\le d(e^t), &&\forall e^t\in E_{\mathrm{com}}^T.\label{eq:dw-local-link}
\end{align}
\end{subequations}

These bounds follow from the shared-resource constraints because all resource usages are nonnegative. They use the original capacities, not the residual capacities left by the current master solution; competition among different requests remains in the master problem.

Let $P_r$ index all admitted deployments defined above, and introduce a zero-resource rejection column $p_{\emptyset}$. The complete column set is $\hat P_r=P_r\cup\{p_{\emptyset}\}$, and the associated finite binary set is denoted by $\mathcal F_r$.\footnote{The degree and flow-balance constraints can allow unused circulations. Deleting them preserves feasibility and cannot increase resource consumption. Thus, representing rejection by the all-zero vector and storing columns without unused circulations entails no loss of optimality.} A column records the execution location of every VNF and the communication and storage links used by every virtual link; it is not merely an ordinary physical routing path. For $p\in\hat P_r$, define the fixed binary coefficients $a_p^r$, $b_{n^t}^{p,v_{r,k}}$, and $g_{e^t}^{p,h_{r,l}}$ as the values of $x_r$, $y_{n^t}^{v_{r,k}}$, and $z_{e^t}^{h_{r,l}}$ in that column. In particular, $a_p^r=1$ for $p\in P_r$, whereas all coefficients of $p_{\emptyset}$ are zero.

Since $\mathcal F_r$ is a finite binary set, $\operatorname{conv}(\mathcal F_r)$ is a bounded polytope. The Minkowski--Weyl theorem \cite{conforti2010polyhedral} gives its representation as convex combinations of finitely many extreme points; here, the distinct binary deployment vectors themselves are those extreme points. Accordingly, every point in $\operatorname{conv}(\mathcal F_r)$ admits the following representation, with the node, VNF, link, and virtual-link indices ranging over their original sets:

\begin{equation}\label{eq:dw-convex-representation}
\begin{aligned}
x_r &= \sum_{p\in\hat P_r} a_p^r w_p^r,\\
y_{n^t}^{v_{r,k}} &= \sum_{p\in\hat P_r} b_{n^t}^{p,v_{r,k}}w_p^r,\\
z_{e^t}^{h_{r,l}} &= \sum_{p\in\hat P_r} g_{e^t}^{p,h_{r,l}}w_p^r,\\
\sum_{p\in\hat P_r}w_p^r &= 1,\qquad w_p^r\ge0.
\end{aligned}
\end{equation}

Continuous weights in \eqref{eq:dw-convex-representation} describe a convex relaxation, not an implementable mixture of deployments. Requiring $w_p^r\in\{0,1\}$ instead selects exactly one complete deployment or rejection for each request.

\subsubsection{Master Problem and Restricted Relaxation}

The computing and communication resource usages of column $p\in\hat P_r$ are respectively denoted by

\begin{equation}\label{eq:dw-column-resources}
\begin{aligned}
\alpha_{n^t}^{r,p} &= \sum_{v_{r,k}\in\hat V_r} f_r\delta_{v_{r,k}}b_{n^t}^{p,v_{r,k}}, &&n^t\in N^T,\\
\beta_{e^t}^{r,p} &= \sum_{h_{r,l}\in H_r} f_r g_{e^t}^{p,h_{r,l}}, &&e^t\in E_{\mathrm{com}}^T.
\end{aligned}
\end{equation}

For compactness, let $\mathcal C=\sum_{n^t\in N^T}c(n^t)$ and $\mathcal D=\sum_{e^t\in E_{\mathrm{com}}^T}d(e^t)$ denote the positive, finite normalization constants in \eqref{eq:sm-node-cost} and \eqref{eq:sm-link-cost}. Each time-expanded node or communication link is counted once. The normalized cost of column $p$ is

\begin{equation}\label{eq:dw-column-cost}
C_p^r=\frac{\psi}{\mathcal C}\sum_{n^t\in N^T}\alpha_{n^t}^{r,p}+\frac{1-\psi}{\mathcal D}\sum_{e^t\in E_{\mathrm{com}}^T}\beta_{e^t}^{r,p}.
\end{equation}

Thus, its objective coefficient is $Ma_p^r-C_p^r$, and both its resource usages and objective coefficient are zero for $p_{\emptyset}$. Storage links remain part of each complete deployment but do not introduce finite-capacity rows or communication-cost terms under the current storage assumption. Substituting the column representation into the original objective and shared-resource constraints yields the integer master problem (MP):

\begin{subequations}\label{eq:dw-master}
\begin{align}
\text{(MP)}\quad &\max_{\mathbf w}\ \sum_{r\in R}\sum_{p\in\hat P_r}(Ma_p^r-C_p^r)w_p^r\label{eq:dw-master-objective}\\
\text{s.t.}\quad &\sum_{p\in\hat P_r}w_p^r=1, &&\forall r\in R,\label{eq:dw-master-convexity}\\
&\sum_{r\in R}\sum_{p\in\hat P_r}\alpha_{n^t}^{r,p}w_p^r\le c(n^t), &&\forall n^t\in N^T,\label{eq:dw-master-node}\\
&\sum_{r\in R}\sum_{p\in\hat P_r}\beta_{e^t}^{r,p}w_p^r\le d(e^t), &&\forall e^t\in E_{\mathrm{com}}^T,\label{eq:dw-master-link}\\
&w_p^r\in\{0,1\}, &&\forall r\in R,\ p\in\hat P_r.\label{eq:dw-master-domain}
\end{align}
\end{subequations}

The full integer MP has the same optimal value as the original formulation. Indeed, a feasible integer column selection recovers request-feasible binary variables through \eqref{eq:dw-convex-representation}, while \eqref{eq:dw-master-node} and \eqref{eq:dw-master-link} enforce their combined resource usages. Conversely, any original feasible solution can be encoded by one column per request after deleting unused circulations, without decreasing its objective value.

Enumerating $\hat P_r$ is generally impractical. We therefore maintain a generated subset $\hat P'_r\subseteq\hat P_r$ containing $p_{\emptyset}$ for every request. Replacing $\hat P_r$ by $\hat P'_r$ in \eqref{eq:dw-master} defines the integer restricted master problem (RMP). During column generation, we solve its linear programming (LP) relaxation, denoted by LP-RMP, by replacing \eqref{eq:dw-master-domain} with $w_p^r\ge0$. The convexity constraints already imply $w_p^r\le1$. Including rejection columns makes this relaxation feasible even when no admitted deployment has yet been generated for some requests. Dual prices are obtained from this LP relaxation, not from the integer RMP \cite{desrosiers2005primer}.

\subsubsection{Request-Wise Pricing}

Let $\sigma_r$ be the dual variable associated with \eqref{eq:dw-master-convexity}, and let $\pi_{n^t}\ge0$ and $\eta_{e^t}\ge0$ correspond to \eqref{eq:dw-master-node} and \eqref{eq:dw-master-link}, respectively, in LP-RMP. The dual problem (DP) is

\begin{equation}\label{eq:dw-rmp-dual}
\begin{aligned}
\text{(DP)}\quad
&\min_{\boldsymbol{\sigma},\boldsymbol{\pi},\boldsymbol{\eta}}\quad
\sum_{r\in R}\sigma_r+\sum_{n^t\in N^T}\pi_{n^t}c(n^t)\\
&\quad+\sum_{e^t\in E_{\mathrm{com}}^T}\eta_{e^t}d(e^t)\\
\text{s.t.}\quad
&\sigma_r+\sum_{n^t\in N^T}\pi_{n^t}\alpha_{n^t}^{r,p}\\
&+\sum_{e^t\in E_{\mathrm{com}}^T}\eta_{e^t}\beta_{e^t}^{r,p}
\ge Ma_p^r-C_p^r,\\
&\qquad\forall r\in R,\ p\in\hat P'_r,\\
&\sigma_r\in\mathbb R,\qquad\forall r\in R,\\
&\pi_{n^t}\ge0,\qquad\forall n^t\in N^T,\\
&\eta_{e^t}\ge0,\qquad\forall e^t\in E_{\mathrm{com}}^T.
\end{aligned}
\end{equation}

Although $\sigma_r$ is the multiplier of an equality constraint and is therefore unrestricted in sign, the dual constraint corresponding to the rejection column implies $\sigma_r\ge0$ at dual feasibility. The dual constraints in \eqref{eq:dw-rmp-dual} cover only the generated columns. Pricing searches for an ungenerated feasible deployment that violates its corresponding dual constraint \cite{desrosiers2005primer}. Accordingly, for the maximization objective, the reduced cost of a column $p\in\hat P_r$ is

\begin{equation}\label{eq:dw-reduced-cost}
\begin{aligned}
\zeta_p^r={}&Ma_p^r-C_p^r-\sigma_r\\
&-\sum_{n^t\in N^T}\pi_{n^t}\alpha_{n^t}^{r,p}-\sum_{e^t\in E_{\mathrm{com}}^T}\eta_{e^t}\beta_{e^t}^{r,p}.
\end{aligned}
\end{equation}

An ungenerated feasible column with positive reduced cost is eligible for addition. Substituting \eqref{eq:dw-column-resources} and \eqref{eq:dw-column-cost} into \eqref{eq:dw-reduced-cost} gives the pricing subproblem (SP) for request $r$, denoted by $\mathrm{SP}_r$:

\begin{equation}\label{eq:dw-pricing}
\begin{aligned}
& (\mathrm{SP}_r)\quad \zeta_r^{\star}=\max_{(x_r,\mathbf y_r,\mathbf z_r)\in\mathcal F_r}\ \Biggl\{Mx_r-\sigma_r\\
&\quad-\sum_{n^t\in N^T}\left(\pi_{n^t}+\frac{\psi}{\mathcal C}\right)\sum_{v_{r,k}\in\hat V_r}f_r\delta_{v_{r,k}}y_{n^t}^{v_{r,k}}\\
&\quad-\sum_{e^t\in E_{\mathrm{com}}^T}\left(\eta_{e^t}+\frac{1-\psi}{\mathcal D}\right)\sum_{h_{r,l}\in H_r}f_rz_{e^t}^{h_{r,l}}\Biggr\}.
\end{aligned}
\end{equation}

The objective augments the original resource costs with prices reflecting shared-resource competition. Thus, pricing is a single-request deployment problem under modified resource costs, and different requests can be priced independently for a fixed set of master dual prices. The rejection column has reduced cost $-\sigma_r\le0$ and is already present, so a search for new admitted columns may equivalently fix $x_r=1$.

Different solution procedures can be used through the same pricing interface. However, an unsuccessful heuristic search or an early-terminated solver run does not by itself establish that no improving column exists.

If LP-RMP is solved to optimality and every pricing subproblem is certified to have $\zeta_r^{\star}\le0$, the current dual solution satisfies the constraints associated with all columns in the full master problem. Meanwhile, the current primal solution remains feasible for the full master LP by assigning zero weights to ungenerated columns. Strong duality then establishes that the current LP-RMP objective equals the optimum of the full master LP relaxation \cite{desrosiers2005primer}.

The full master LP optimum provides an upper bound on the original maximization problem. Before pricing completeness is certified, the LP-RMP objective alone need not provide such an upper bound, because potentially useful columns remain excluded. At any configured termination, a feasible integer selection from the generated pool provides a feasible deployment and hence a lower bound on the original optimum. Even if this finite-pool integer problem is solved optimally, global integer optimality is generally not certified without an additional valid bound or a complete branch-and-price procedure \cite{barnhart1998branch,vanderbeck2000dantzig}.

\subsection{SAGINFE: Fast Feasible Construction}\label{subsec:saginfe}

We develop SAGIN Flow Embedding (SAGINFE) to efficiently construct a feasible solution and generate initial columns for the subsequent joint optimization. SAGINFE coordinates multiple requests under shared residual resource constraints, while its Per-Flow Embedding (PFE) procedure constructs a complete deployment for each individual request. PFE follows a pseudo-dynamic-programming design: it advances through successive VNF stages using a cost recursion inspired by dynamic programming \cite{bellman1954theory}, but retains only one deployment record for each candidate location.

\subsubsection{Multi-Request Coordination}

Algorithm~\ref{alg:saginfe} presents SAGINFE. Let $\bar c(n^t)$ and $\bar d(e^t)$ denote the residual computing capacity of node $n^t\in N^T$ and the residual data capacity of communication link $e^t\in E_{\mathrm{com}}^T$, respectively. They are initialized to $c(n^t)$ and $d(e^t)$. SAGINFE sorts the requests in ascending order of $(|V_r|,f_r)$, giving priority to shorter SFCs and then smaller data amounts; exact ties preserve the input order. It is a greedy admission rule rather than an optimal request ordering.

\begin{algorithm}[t]
\small
\caption{SAGINFE}\label{alg:saginfe}
\LinesNumbered
\KwIn{TEG $G=(N^T,E^T)$, capacities $c,d$, requests $R$.}
\KwOut{A feasible deployment and initial column sets $\hat P'_r$.}
Initialize $\bar c\leftarrow c$, $\bar d\leftarrow d$ on communication links\;
Initialize $x_r\leftarrow0$, $\mathbf y_r\leftarrow\mathbf0$, $\mathbf z_r\leftarrow\mathbf0$, and $\hat P'_r\leftarrow\{p_{\emptyset}\}$ for all $r\in R$\;
Sort requests by ascending $(|V_r|,f_r)$\;
\For{each request $r$ in the sorted order}{
  Call PFE$(G,r,\bar c,\bar d)$ using temporary resource records\;
  \If{PFE returns a complete feasible deployment}{
    Set $x_r\leftarrow1$ and retain the returned $\mathbf y_r,\mathbf z_r$\;
    Commit the returned residual capacities to $\bar c,\bar d$\;
    Encode the complete deployment as $p_r^{\mathrm S}$ and calculate its resource usages using \eqref{eq:dw-column-resources}\;
    $\hat P'_r\leftarrow\hat P'_r\cup\{p_r^{\mathrm S}\}$\;
  }
}
Evaluate the solution using \eqref{eq:sm-objective}\;
\Return{the feasible solution and $\{\hat P'_r:r\in R\}$}\;
\end{algorithm}

\subsubsection{Stage-Wise Per-Flow Embedding}

PFE constructs $|V_r|+2$ stages, indexed by $k=0,1,\ldots,|V_r|+1$, corresponding to the extended VNF sequence $\hat V_r$. Define the time-expanded node set within the service window as $N_r^T=\{n^t\in N^T:t_{r,o}\le t\le t_{r,d}^{\max}\}$. The candidate locations at stage $k$ are

\begin{equation}\label{eq:pfe-candidates}
A_{r,k}=\begin{cases}
\{n_{r,o}^{t_{r,o}}\}, & k=0,\\
N_r^T, & 1\le k\le |V_r|,\\
\{n_{r,d}^{t}:t_{r,o}\le t\le t_{r,d}^{\max}\}, & k=|V_r|+1.
\end{cases}
\end{equation}

\begin{figure}[t]
\captionsetup{font={small}}
\centering
\includegraphics[width=1\columnwidth,height=0.2\textheight,keepaspectratio]{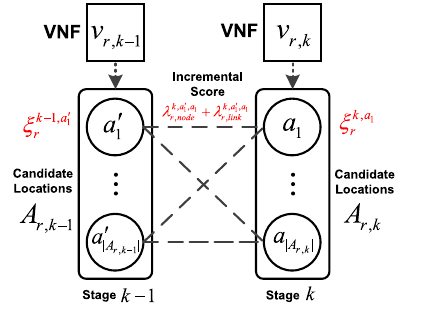}
\caption{Illustration of two consecutive VNF stages in PFE. Each inter-stage connection represents a TEG route. For each next-stage candidate, PFE retains the feasible extension with the lowest accumulated heuristic score.}\label{fig:pfe_illustration}
\end{figure}

The construction is illustrated in Fig.~\ref{fig:pfe_illustration}; an inter-stage connection represents a complete TEG route, potentially including several communication and storage links.

Each retained candidate $a\in A_{r,k}$ is associated with the record $\mathcal L_r^{k,a}=\bigl(\mathbf c_r^{k,a},\mathbf D_r^{k,a},\mathbf y_r^{k,a},\mathbf z_r^{k,a},\xi_r^{k,a}\bigr)$. Here, $\mathbf c_r^{k,a}$ and $\mathbf D_r^{k,a}$ record the residual node and communication-link capacities after the partial deployment ending at $a$; their entries are denoted by $c_r^{k,a}(n^t)$ and $d_r^{k,a}(e^t)$. The vectors $\mathbf y_r^{k,a}$ and $\mathbf z_r^{k,a}$ record the VNF locations through stage $k$ and the routes of the corresponding virtual links, respectively. The scalar $\xi_r^{k,a}$ is the accumulated heuristic score. The source record is initialized from $\bar c,\bar d$, with $v_{r,0}$ fixed at $n_{r,o}^{t_{r,o}}$, no deployed virtual links, and score zero.

To construct stage $k\ge1$, consider a retained predecessor $a'\in A_{r,k-1}$ and a target $a\in A_{r,k}$. The incremental computing score is

\begin{equation}\label{eq:pfe-node-score}
\lambda_{r,\mathrm{node}}^{k,a',a}=\begin{cases}
0, & f_r\delta_{v_{r,k}}=0,\\[2pt]
\dfrac{\psi f_r\delta_{v_{r,k}}}{c_r^{k-1,a'}(a)}, & 0<f_r\delta_{v_{r,k}}\le c_r^{k-1,a'}(a),\\[5pt]
\infty, & \text{otherwise}.
\end{cases}
\end{equation}

This score charges a larger fraction when the same computation uses a more depleted node. The zero-demand case also covers the destination dummy VNF and avoids division by zero at a node with no computing capacity.

Routing searches the full TEG with targets restricted to $A_{r,k}$. For predecessor record $\mathcal L_r^{k-1,a'}$, remove unavailable communication links and those with $d_r^{k-1,a'}(e^t)<f_r$. Each remaining link receives weight

\begin{equation}\label{eq:pfe-edge-score}
u_r^{k,a'}(e^t)=\begin{cases}
\dfrac{(1-\psi)f_r}{d_r^{k-1,a'}(e^t)}, & e^t\in E_{\mathrm{com}}^T,\\[5pt]
0, & e^t\in E_{\mathrm{sto}}^T.
\end{cases}
\end{equation}

Since all weights are nonnegative, PFE runs a single-source Dijkstra search \cite{dijkstra1959note} from $a'$ and reuses its distances for all targets in $A_{r,k}$. Let $\varphi_r^{k,a',a}$ be the resulting route for virtual link $h_{r,k}$, with routing score $\lambda_{r,\mathrm{link}}^{k,a',a}=\sum_{e^t\in\varphi_r^{k,a',a}}u_r^{k,a'}(e^t)$.

The routing score is set to $\infty$ if no route exists. When $a'=a$, the empty route has zero score, allowing consecutive VNFs to execute at the same node in the same slot.

PFE selects the lowest-score extension for each target according to

\begin{equation}\label{eq:pfe-recursion}
\xi_r^{k,a}=\min_{a'\in A_{r,k-1}}\left\{\xi_r^{k-1,a'}+\lambda_{r,\mathrm{node}}^{k,a',a}+\lambda_{r,\mathrm{link}}^{k,a',a}\right\},
\end{equation}

where predecessors without a retained record are excluded. If the minimum is finite, let $a^*$ be a minimizing predecessor. PFE copies its resource records, places $v_{r,k}$ at $a$, and appends route $\varphi_r^{k,a^*,a}$ for $h_{r,k}$. The new residual capacities satisfy

\begin{equation}\label{eq:pfe-resource-update}
\begin{aligned}
c_r^{k,a}(n^t)&=c_r^{k-1,a^*}(n^t)-f_r\delta_{v_{r,k}}\mathbf1\{n^t=a\},\\
d_r^{k,a}(e^t)&=d_r^{k-1,a^*}(e^t)-f_r\mathbf1\{e^t\in\varphi_r^{k,a^*,a}\},
\end{aligned}
\end{equation}

for $n^t\in N^T$ and $e^t\in E_{\mathrm{com}}^T$, respectively. Each candidate receives an independent copy, so one tentative extension does not consume another candidate's resources. If different virtual links reuse a communication link, each use is charged separately, as required by \eqref{eq:sm-link-capacity}. Algorithm~\ref{alg:pfe} summarizes this procedure.

\begin{algorithm}[t]
\small
\caption{Per-Flow Embedding (PFE)}\label{alg:pfe}
\LinesNumbered
\KwIn{TEG $G$, request $r$, shared residual capacities $\bar c,\bar d$.}
\KwOut{A complete deployment and updated capacities, or failure.}
Construct $A_{r,k}$ using \eqref{eq:pfe-candidates}\;
Initialize the source record from $\bar c,\bar d$ with score zero\;
\For{$k=1,\ldots,|V_r|+1$}{
  Set $\xi_r^{k,a}\leftarrow\infty$ and clear the selected predecessor for each $a\in A_{r,k}$\;
  \For{each retained predecessor $a'\in A_{r,k-1}$}{
    Build its feasible routing graph and weights using \eqref{eq:pfe-edge-score}\;
    Run single-source Dijkstra from $a'$\;
    \For{each reachable $a\in A_{r,k}$}{
      Calculate $\lambda_{r,\mathrm{node}}^{k,a',a}$ and $\lambda_{r,\mathrm{link}}^{k,a',a}$\;
      Evaluate the candidate score in \eqref{eq:pfe-recursion}\;
      \If{this score is smaller than $\xi_r^{k,a}$}{
        Update $\xi_r^{k,a}$ and save the predecessor and route information\;
      }
    }
  }
  \If{no target has a finite score}{
    \Return{failure; keep the input $\bar c,\bar d$ unchanged}\;
  }
  \For{each target $a$ with a finite score}{
    Copy the selected predecessor's residual records\;
    Extend the placement and routing records\;
    Update capacities using \eqref{eq:pfe-resource-update} and retain $\mathcal L_r^{k,a}$\;
  }
}
Select a final-stage record with the smallest score\;
Trace its predecessors to recover $\mathbf y_r,\mathbf z_r$\;
\Return{the deployment and its final residual capacities}\;
\end{algorithm}

\subsubsection{Feasibility and Initial Columns}

Every returned deployment respects the VNF order because stage $k$ extends a deployment ending at stage $k-1$ through the route of $h_{r,k}$. The candidate sets enforce the endpoints and deadline, while the TEG enforces temporal progression. The capacity tests and independent resource updates maintain nonnegative residual capacities after every extension, including repeated use of a resource by the same request. Applying the same updates across admitted requests therefore produces a jointly feasible solution. However, two partial deployments ending at the same $(k,a)$ can leave different residual resources. Retaining only the one with the smallest current score may discard a deployment that would enable better later extensions \cite{irnich2005resource}. Thus, PFE is a pseudo-dynamic-programming heuristic rather than an exact dynamic program; failure does not certify that the request is infeasible.

For the computational complexity, let $n_T=|N^T|$ and $m_T=|E^T|$. With sparse edge records and a binary heap, each predecessor requires one Dijkstra search taking \mbox{$O((n_T+m_T)\log(1+n_T))$} time. At most $n_T$ predecessors are examined per stage, and copying the resource records for the retained targets takes $O(n_T(n_T+m_T))$ time per stage. A score improvement can trigger an $O(n_T)$ path reconstruction for each of at most $n_T^2$ predecessor--target pairs per stage. Thus, a conservative bound on the PFE runtime is $O\!\left((|V_r|+1)[n_T(n_T+m_T)\log(1+n_T)+n_T^3]\right)$. SAGINFE adds $O(|R|\log(1+|R|))$ sorting time and one PFE call per request.

For each admitted request, its complete deployment $p_r^{\mathrm S}\in P_r$ is inserted into $\hat P'_r$ together with $p_{\emptyset}$; a rejected request initially has only $p_{\emptyset}$. Selecting $p_r^{\mathrm S}$ for admitted requests and $p_{\emptyset}$ for the others reproduces the SAGINFE solution as a feasible integer RMP solution.

\subsection{Configurable Pricing and Parallel Execution}\label{subsec:configurable-pricing}

The illustrated solver-first workflow uses primary solver pricing followed by optional selective PFE. Alternative configurations may use heuristic-first pricing, in which heuristic search precedes strict solver fallback.

\subsubsection{Pricing Schedules}

After an optimal LP-RMP solution is obtained, its dual prices $(\boldsymbol\sigma,\boldsymbol\pi,\boldsymbol\eta)$ are fixed for the next pricing phase. Solver-based pricing uses \eqref{eq:dw-pricing} in two modes. The early-terminated mode limits the number of feasible solutions explored and relaxes the solver's optimality tolerance to obtain candidates quickly. The strict mode removes this early-termination restriction and seeks a more thoroughly optimized pricing solution, subject to the configured computation budget.

An early-terminated-only schedule uses the first mode throughout, whereas a strict-only schedule uses the second. A hybrid schedule starts with early-terminated pricing and switches to strict pricing after a configured number of rounds, or earlier if an early phase produces no eligible new column and sufficient budget remains. Each schedule can be used with or without selective PFE.

\subsubsection{Selective PFE and Candidate Checks}

When enabled, selective PFE calls supplement solver-generated candidates after a primary pricing phase. Requests for which primary pricing has already supplied a new column are excluded from supplementary selection in that phase. Among the remaining requests, priority is given to those represented wholly or partly by rejection or having few admitted-deployment alternatives. The ordering uses the LP rejection weight $w_{p_{\emptyset}}^r$, LP fractionality, and the number of admitted-deployment columns, with request identifiers resolving ties deterministically. Only a limited prefix of this ordering is processed, subject to per-round and overall call limits. This supplementary stage is skipped when initialization already admits all requests or the remaining computation allowance is insufficient after reserving time for integer recovery.

For each selected request, PFE starts from independent copies of the original capacities $c(n^t)$ and $d(e^t)$, rather than the resources left by the SAGINFE solution. The procedure retains the residual-ratio scoring of Section~\ref{subsec:saginfe}; master dual prices enter the subsequent reduced-cost check, not that stage-wise scoring rule. With unchanged inputs and tie-breaking rules, repeated PFE calls may yield the same deployment, which is reassessed through the current reduced cost and duplicate check.

All returned candidates undergo the same feasibility and reduced-cost checks. Their placements and virtual-link routes are checked against the request-specific constraints; single-request resource bounds are enforced by the built-in candidate generators. Their resource usages and normalized costs are recomputed using \eqref{eq:dw-column-resources} and \eqref{eq:dw-column-cost}, and their reduced costs are evaluated with the dual prices of the current phase. A candidate is eligible for insertion only if

\begin{equation}\label{eq:cg-candidate-check}
p\in P_r,\qquad p\notin\hat P'_r,\qquad \zeta_p^r>\varepsilon_{\mathrm{rc}},
\end{equation}

where $\varepsilon_{\mathrm{rc}}>0$ is the numerical reduced-cost tolerance. Duplicate detection compares the complete VNF placement and ordered virtual-link routes, including storage links.

\begin{algorithm}[t]
\small
\caption{Configurable Column Generation with SAGINFE Initialization}\label{alg:configurable-cg}
\LinesNumbered
\KwIn{TEG $G$, requests $R$, pricing schedule, optional PFE settings, worker count, and computation budgets.}
\KwOut{A feasible deployment $\mathcal S_{\mathrm{out}}$.}
Obtain $\mathcal S_0$ and initial column sets $\hat P'_r$ using SAGINFE\;
\While{the configured pricing budget permits another round}{
  Solve LP-RMP over the current column sets\;
  \If{no optimal LP solution is available}{\textbf{break}\;}
  Fix the dual prices and select the scheduled pricing mode\;
  Form the ordered solver-pricing tasks and account for their work budget\;
  Execute the tasks sequentially or with the configured workers\;
  Check and insert eligible columns in task order using \eqref{eq:cg-candidate-check}\;
  \If{selective PFE is enabled and its eligibility and budget conditions hold}{
    Select requests and generate PFE candidates\;
    Apply the same candidate checks and insert eligible columns\;
  }
  \If{a complete pricing certificate has been obtained}{\textbf{break}\;}
  \If{no eligible column was added}{
    \eIf{the policy and remaining budget permit a stricter phase}{
      Activate strict pricing for the next round\;
    }{\textbf{break}\;}
  }
}
Fix the final generated column sets\;
\eIf{the most recently solved LP-RMP has an incumbent with numerically integral column weights}{
  Reuse its column selection, assigning zero weights to any later-added columns\;
}{
  Attempt integer RMP recovery over the final column sets within its allowance, optionally using $\mathcal S_0$ as an initial feasible solution\;
}
Validate any recovered pool solution and recompute its objective\;
Choose $\mathcal S_{\mathrm{out}}$ using \eqref{eq:cg-solution-safeguard}, or retain $\mathcal S_0$ if no valid pool solution is available\;
\Return{$\mathcal S_{\mathrm{out}}$}\;
\end{algorithm}

\subsubsection{Request-Wise Parallel Execution}

For fixed master dual prices, solver-pricing tasks for different requests can run independently. Before dispatch, the coordinator forms an ordered task list and accounts for its calls against the pricing budget. Each worker receives one request, the common dual prices, and fixed network data, and maintains its own solver state. The worker count controls concurrency; each pricing solver uses one thread in the parallel configuration.

The coordinator collects results and processes them in the predefined task order through the checks in \eqref{eq:cg-candidate-check}. Workers do not modify the shared column pool, and the next LP-RMP solve follows completion of the current batch. In the implementation, this parallelism applies to solver-based pricing; SAGINFE initialization, selective PFE, and master optimization remain outside the worker pool.

The pricing loop ends when a complete pricing certificate is obtained, the configured round or work budget is exhausted, or no eligible column is found and no further pricing phase is permitted. A complete certificate requires all requests to be covered under the same fixed dual prices, with optimal pricing values or valid upper bounds establishing that no positive-reduced-cost column remains. A strict solver setting alone does not establish this condition if a solve terminates before obtaining the required bound. Numerically, a test that allows reduced costs up to $\varepsilon_{\mathrm{rc}}$ establishes termination within that tolerance, rather than the exact zero-threshold condition in Section~\ref{subsec:dw-decomposition}.

When an overall time limit is imposed, part of it is reserved for integer recovery. Algorithm~\ref{alg:configurable-cg} summarizes the SAGINFE-initialized, solver-first workflow, including the final selection described next.

\subsection{Finite-Pool Integer Selection and Solution Safeguard}\label{subsec:integer-selection}

At the end of pricing, the generated sets $\hat P'_r$ are fixed. If the most recently solved LP-RMP has an incumbent with numerically integral column weights, its column selection can be reused after validation. Otherwise, we attempt integer RMP recovery by restricting \eqref{eq:dw-master} to these final sets. A reused LP selection may precede the last column additions; assigning zero weights to the newly added columns preserves feasibility. Direct reuse does not by itself certify optimality over the final pool. The selected columns are converted back to the original variables through \eqref{eq:dw-convex-representation}, and the resulting request deployments and aggregate resource usages are validated.

When integer RMP recovery is invoked after pricing, the retained initial and rejection columns allow the SAGINFE selection in Section~\ref{subsec:saginfe} to be supplied as an initial feasible solution to the solver. Integer optimization has a separate time allowance and may return a feasible incumbent before proving restricted-pool optimality.

Let $\mathcal S_0$ denote the validated SAGINFE solution and $\mathcal S_{\mathrm{pool}}$ a validated integer solution obtained from the generated pool. Write $\mathcal J(\mathcal S)$ for a solution's objective value in \eqref{eq:sm-objective}. When $\mathcal S_{\mathrm{pool}}$ is available, the framework returns

\begin{equation}\label{eq:cg-solution-safeguard}
\mathcal S_{\mathrm{out}}\in
\operatorname*{arg\,max}_{\mathcal S\in\{\mathcal S_0,\mathcal S_{\mathrm{pool}}\}}
\mathcal J(\mathcal S).
\end{equation}

Otherwise, it returns $\mathcal S_0$. Consequently, $\mathcal J(\mathcal S_{\mathrm{out}})\ge\mathcal J(\mathcal S_0)$ under the admission-first objective: the returned solution admits at least as many requests as SAGINFE and has no higher cost when the counts are equal. This guarantee holds independently of whether the finite-pool integer problem is solved to optimality.

The final integer search operates on the generated pool without further pricing inside its branching tree \cite{joncour2010primal}.

The preceding configuration choices yield different deployment algorithms within the same framework. DWCG-Pure (Pure) starts from rejection columns and uses solver pricing without PFE. DWCG-SAGINFE (HFP, denoting heuristic-first pricing) combines SAGINFE initialization with heuristic search followed by strict solver fallback. DWCG-SAGINFE-PRO (PRO) adopts the SAGINFE-initialized, solver-first workflow in Algorithm~\ref{alg:configurable-cg}, combining selective PFE with finite-pool integer selection, an initial feasible solution for the integer RMP solver, and the solution safeguard. We adopt PRO as the default configuration for its favorable overall admission--runtime balance in the experiments of Section~\ref{sec:simulation-results}.

\section{Simulation Results}\label{sec:simulation-results}

In this section, we evaluate the admission performance and computational efficiency of the proposed framework.

\subsection{Simulation Setup}\label{subsec:simulation-setup}

The algorithms are implemented in Python 3.8.10 with Gurobi 11.0.3~\cite{gurobioptimizer} on a mobile workstation equipped with an Intel Core i7-12700H processor (14 cores and 20 logical processors) and 32~GiB of RAM\footnote{Source code for all algorithms and simulations, together with detailed experimental settings and results, is publicly available at \url{https://github.com/flgkd/SAGIN-SFC-deployment-heuristic-assisted-CG}.}. Each solver instance uses one thread. Unless otherwise specified, pricing is sequential; the parallel experiments use one, two, or four worker processes, each running a single-threaded solver.

\subsubsection{Network and Request Generation}

We generate synthetic SAGINs at four scales, denoted by S20, S30, S40, and S50, containing $(8,4,8)$, $(12,6,12)$, $(16,8,16)$, and $(20,10,20)$ space, air, and ground nodes, respectively. Within each layer, nodes are indexed from zero as $n_{s_i}$, $n_{a_j}$, and $n_{g_\ell}$. Ground nodes are divided into four equal regions of consecutive indices. With $q=|N_g|/4$, ground node $n_{g_\ell}$ has region index $\rho=\lfloor\ell/q\rfloor$ and within-region position $k=\ell\bmod q$. For all four scales, $|N_s|=2|N_a|$.

Table~\ref{tab:simulation-topology-rules} specifies all connections in the seed-independent undirected candidate graph. There are no direct space--ground or inter-region ground--ground connections. For S20, these rules yield 48 physical connections: 16 SS, 8 SA, 4 AA, 16 AG, and 4 GG.

\begin{table}[t]
\small
\captionsetup{font={footnotesize}}
\caption{Candidate Topology Generation Rules}\label{tab:simulation-topology-rules}
\centering
\setlength{\tabcolsep}{4pt}
\renewcommand{\arraystretch}{1.08}
\begin{tabular}{|>{\centering\arraybackslash}p{0.10\columnwidth}|>{\raggedright\arraybackslash}p{\dimexpr0.90\columnwidth-4\tabcolsep-3\arrayrulewidth\relax}|}
\hline
\textbf{Type} & \centering\arraybackslash\textbf{Connection Rule} \\
\hline
SS & Each space node $i$ connects to space nodes $(i+1)\bmod |N_s|$ and $(i+2)\bmod |N_s|$, forming a ring with second-neighbor links. \\
\hline
SA & Air node $j$ connects to space nodes $2j$ and $(2j+|N_s|/2)\bmod |N_s|$. \\
\hline
AA & Each air node $j$ connects to air node $(j+1)\bmod |N_a|$, forming a ring. \\
\hline
AG & Ground node $\ell$ connects to air nodes $\rho$ (primary) and $(\rho+1+k)\bmod |N_a|$ (secondary). \\
\hline
GG & Ground nodes within each region form a ring in increasing index order; a two-node region contains only one connection. \\
\hline
\end{tabular}
\par\smallskip
\begin{minipage}{\columnwidth}
\footnotesize S, A, and G denote space, air, and ground. Node indices are local to each layer. Rules apply to every node in the indicated layer or region; connections are undirected, and duplicate unordered node pairs are included only once.
\end{minipage}
\end{table}

For each time slot, a new snapshot is sampled from the same candidate graph, independently of previous snapshots. Each candidate connection is independently retained using the type-dependent probabilities in Table~\ref{tab:simulation-parameters}. Disconnected snapshots are discarded in full and resampled, with at most 100 attempts; generation fails if no connected snapshot is obtained. These probabilities therefore apply before conditioning on connectivity. All physical nodes are retained in every snapshot. The complete sequence of snapshots is generated before planning and provided as known input.

\begin{table}[t]
\small
\captionsetup{font={footnotesize}}
\caption{Resource and Request Parameters}\label{tab:simulation-parameters}
\centering
\setlength{\tabcolsep}{4pt}
\renewcommand{\arraystretch}{1.08}
\begin{tabular}{|>{\raggedright\arraybackslash}p{0.43\columnwidth}|>{\raggedright\arraybackslash}p{\dimexpr0.57\columnwidth-4\tabcolsep-3\arrayrulewidth\relax}|}
\hline
\centering\textbf{Parameter} & \centering\arraybackslash\textbf{Setting} \\
\hline
Base node capacity & \mbox{Space: 64--96}; \mbox{air: 24--48}; \mbox{ground: 80--120} \\
\hline
Base communication capacity & \mbox{SS: 64--96}; \mbox{SA: 28--56}; \mbox{AA: 16--40}; \mbox{AG: 21--42}; \mbox{GG: 64--120} \\
\hline
Link retention probability & \mbox{SS: 0.95}; \mbox{SA: 0.85}; \mbox{AA: 0.80}; \mbox{AG: 0.80}; \mbox{GG: 0.95} \\
\hline
Slot-wise capacity multiplier & Uniform on $[0.9,1.1]$ \\
\hline
Requests $|R|$ & 80,120,160,200 for S20--S50 \\
\hline
Base request data amount & Uniform on $\{12,\ldots,24\}$ \\
\hline
Load multiplier & \mbox{Low: 0.4}; \mbox{Medium: 0.6}; \mbox{High: 0.8} \\
\hline
Flow length $|\hat V_r|$ & 4,5,6 with proportions 0.2,0.5,0.3 \\
\hline
Processing coefficient $\delta_{v_{r,k}}$ & Uniform on $[2.5,5.0]$ \\
\hline
Default deadline offset $\Delta_r$ & 2,3,4 slots with probabilities 0.3,0.4,0.3 \\
\hline
Objective parameters & $M=2$, $\psi=0.5$ \\
\hline
\end{tabular}
\par\smallskip
\begin{minipage}{\columnwidth}
\footnotesize S, A, and G denote space, air, and ground, respectively. SS, SA, AA, AG, and GG identify link types. Base capacities are uniformly sampled from inclusive integer ranges; communication capacities are data amounts per slot.
\end{minipage}
\end{table}

\begin{table}[t]
\small
\captionsetup{font={footnotesize}}
\caption{Methods in the Main Comparison}\label{tab:simulation-methods}
\centering
\setlength{\tabcolsep}{4pt}
\renewcommand{\arraystretch}{1.08}
\begin{tabular}{|>{\raggedright\arraybackslash}p{0.25\columnwidth}|>{\raggedright\arraybackslash}p{\dimexpr0.75\columnwidth-4\tabcolsep-3\arrayrulewidth\relax}|}
\hline
\centering\textbf{Method} & \centering\arraybackslash\textbf{Configuration} \\
\hline
SAGINFE & Standalone feasible construction without CG. \\
\hline
Pure & Rejection-only initial pool; solver pricing; no PFE. \\
\hline
HFP & SAGINFE initialization; heuristic-first pricing with strict solver fallback. \\
\hline
PRO & SAGINFE initialization; solver pricing followed by selective PFE; SAGINFE solution supplied to the integer RMP solver; solution safeguard. \\
\hline
\end{tabular}
\par\smallskip
\begin{minipage}{\columnwidth}
\footnotesize Pure and PRO use HYB-S5 by default. HFP invokes heuristic pricing before solver fallback, whereas PRO applies selective PFE after primary solver pricing.
\end{minipage}
\end{table}

Base capacities are sampled uniformly from the inclusive integer ranges in Table~\ref{tab:simulation-parameters}, then independently multiplied by a continuous uniform factor in $[0.9,1.1]$ for each node or link in each slot. An active bidirectional connection becomes two directed communication links with equal, separately accounted capacities. Their values specify $d(e^t)$ in data units per slot. Consecutive copies of the same node are connected by storage links with unlimited capacity and zero communication cost. These synthetic resource ranges provide controlled heterogeneity across network layers; capacities and demands use consistent model units. The default horizon contains eight slots.

Request endpoints comprise 70\% cross-region ground--ground pairs, 20\% same-region ground--ground pairs, and 10\% ground--nonground pairs. Endpoint categories and flow lengths follow the prescribed proportions using integer quotas. VNF types are sampled with replacement from a catalogue containing firewall, network address translation, and deep packet inspection~\cite{halpern2015sfc}, together with intrusion detection, caching, and transcoding. Each processing stage independently receives a coefficient $\delta_{v_{r,k}}$ from Table~\ref{tab:simulation-parameters}. The data amount $f_r$ equals the sampled base amount multiplied by the load factor. Under the default Balanced deadline profile, the offset $\Delta_r=t_{r,d}^{\max}-t_{r,o}$ follows Table~\ref{tab:simulation-parameters}; $t_{r,o}$ is drawn uniformly from $\{1,\ldots,|T|-\Delta_r\}$, and the deadline is inclusive.

\subsubsection{Compared Methods and Configurations}

The main comparison includes SAGINFE, DWCG-Pure (Pure), DWCG-SAGINFE (HFP, denoting heuristic-first pricing), and DWCG-SAGINFE-PRO (PRO), whose configurations are summarized in Table~\ref{tab:simulation-methods}. Pure adapts the solver-based column-generation approach of Jia et al.~\cite{jia2020virtual} to our deployment model and serves as a baseline for evaluating the added components, while PRO is the default configuration for its favorable overall admission--runtime balance demonstrated in the following experiments.

\begin{figure}[t]
\captionsetup{font={small}}
\centering
\includegraphics[width=1\columnwidth,height=0.8\textheight,keepaspectratio]{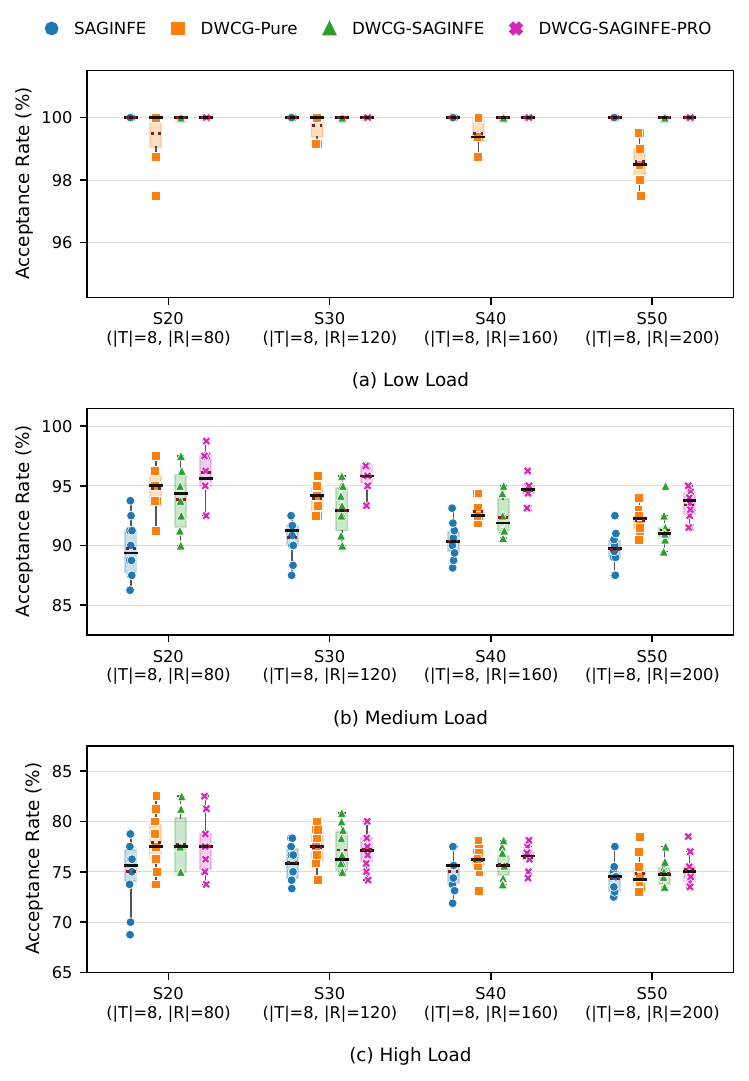}
\caption{Admission ratios across network scales under Low, Medium, and High load, with ten paired instances per setting. The vertical ranges differ across load panels. DWCG-SAGINFE in the legend denotes HFP.}\label{fig:results-main-admission}
\end{figure}

The default solver-pricing schedule for Pure and PRO, denoted by HYB-S5, uses early-terminated pricing for the first five rounds and strict pricing thereafter. Strict pricing sets the solver's relative and absolute optimality gap tolerances to 0 and $10^{-7}$, respectively; early-terminated pricing uses a solution limit of one and a relative optimality gap tolerance of 0.1. CG allows at most 12 rounds and $13|R|$ solver-pricing calls, while final integer selection has a 300~s limit. Its feasible incumbent is used if optimality over the finite pool is not certified. Selective PFE permits at most $\min\{\lceil0.05|R|\rceil,10\}$ calls per round and $\min\{\lceil0.20|R|\rceil,40\}$ calls in total. Candidate columns must pass feasibility and duplicate checks and have recomputed reduced cost greater than $10^{-6}$.

For small-instance references, we solve the original ILP with two objectives. ILP-AdmissionOnly maximizes the admitted count on S20 and S30 at Low and Medium loads; ILP-FullObjective optimizes~\eqref{eq:sm-objective} at Low load. Only certified optima are used as exact references.

\subsubsection{Evaluation Protocol and Metrics}

The main experiments combine all four scales and three load levels. To examine the effect of TEG size, the fixed-flow horizon study uses S40, 160 requests, and Medium load with $|T|\in\{6,8,10,12\}$. It retains the complete request bundle generated for six slots and takes prefixes of a common twelve-slot topology sequence. All deadlines therefore remain within the first six slots.

Each experimental setting is evaluated on ten instances using seeds 0--9, with identical network and request instances shared by the compared methods within that experiment. We report admitted request counts, admission ratios, normalized resource costs, and runtime. Admission ratios equal $100\sum_{r\in R}x_r/|R|$. Runtime excludes instance generation and includes model construction, initialization, master optimization, pricing, worker startup and dispatch, and final integer selection, as applicable. Aggregate comparisons use mean admission and median runtime unless stated otherwise. Cross-method cost comparisons use common full-admission instances to keep the admitted request set identical. In the distribution figures, boxes show the interquartile range, whiskers extend to the most extreme observations within 1.5 times that range, points represent individual seeds, black solid marks indicate medians, and red dotted marks indicate means; these are descriptive summaries, not confidence intervals. Pricing-stage and end-to-end parallel speedups are computed separately as the sequential-to-parallel runtime ratio for each seed, and then summarized across the ten paired instances.

\subsection{Admission Performance and Optimality Gaps}\label{subsec:results-admission}

\subsubsection{Admission Across Scales and Loads}

Fig.~\ref{fig:results-main-admission} compares admission ratios across network scales and loads. At Low load, SAGINFE, HFP, and PRO admit all requests on every instance, whereas Pure occasionally leaves requests unadmitted, with its mean admission ratio decreasing to 98.60\% at S50. At Medium load, PRO achieves the highest mean admission at all four scales: 96.13\%, 95.75\%, 94.56\%, and 93.40\% for S20--S50, respectively. Relative to SAGINFE, these results correspond to gains of 5.1, 6.1, 6.7, and 7.5 admitted requests on average. PRO also admits 1.1--2.7 more requests on average than Pure in these Medium-load cases. At High load, the differences among CG configurations become smaller: Pure has slightly higher mean admission at S20 and S30, while PRO has higher means at S40 and S50. The preferred configuration therefore depends on the operating conditions.

\begin{figure*}[!t]
\centering
\captionsetup{font={small}}

\begin{minipage}[t]{0.5\textwidth}
\centering
\vspace{0pt}
\includegraphics[width=\linewidth]{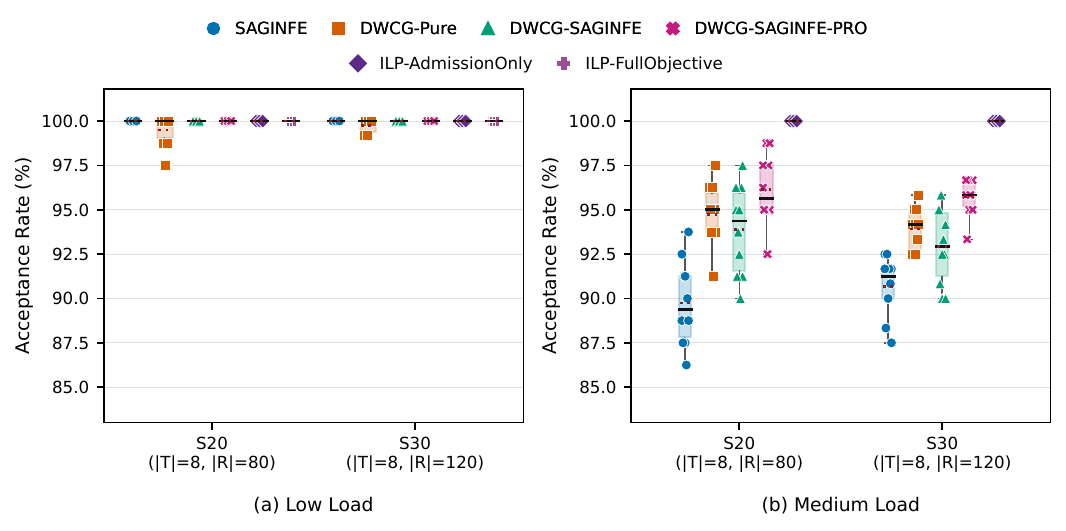}
\par\smallskip
{\small\textbf{(A)}~Admission comparison.\par}
\end{minipage}\hfill
\begin{minipage}[t]{0.5\textwidth}
\centering
\vspace{0pt}
\includegraphics[width=\linewidth]{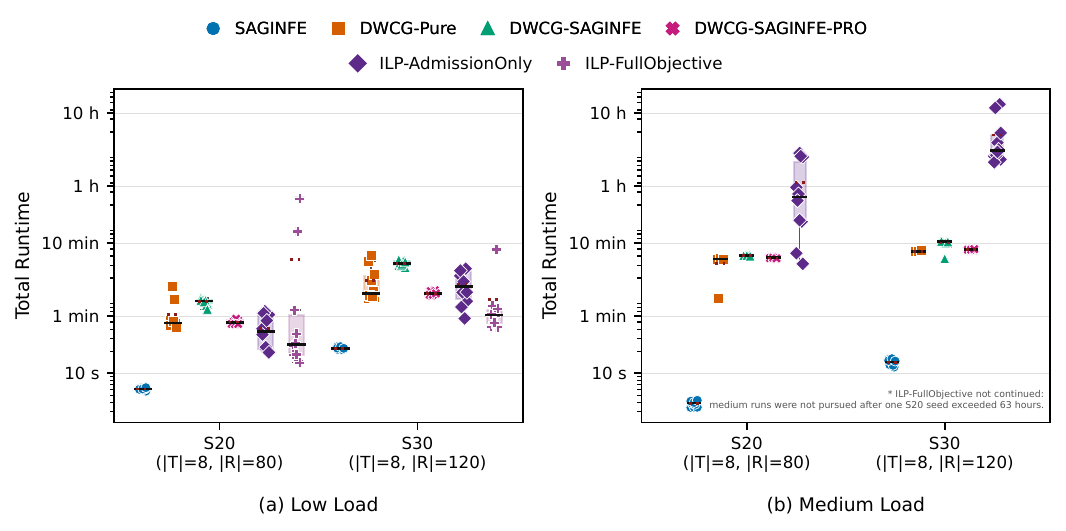}
\par\smallskip
{\small\textbf{(B)}~End-to-end runtime.\par}
\end{minipage}

\par\medskip

\begin{minipage}[t]{0.5\textwidth}
\centering
\vspace{0pt}
\parbox[c][0.62\linewidth][c]{\linewidth}{\centering
\includegraphics[width=0.9\linewidth]{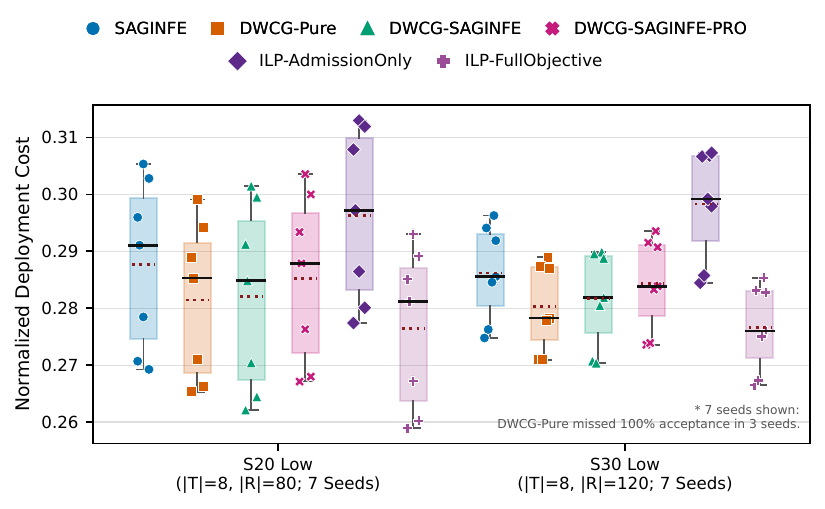}
}
\par\smallskip
{\small\textbf{(C)}~Cost: six methods, seven seeds per scale.\par}
\end{minipage}\hfill
\begin{minipage}[t]{0.5\textwidth}
\centering
\vspace{0pt}
\parbox[c][0.62\linewidth][c]{\linewidth}{\centering
\includegraphics[width=\linewidth]{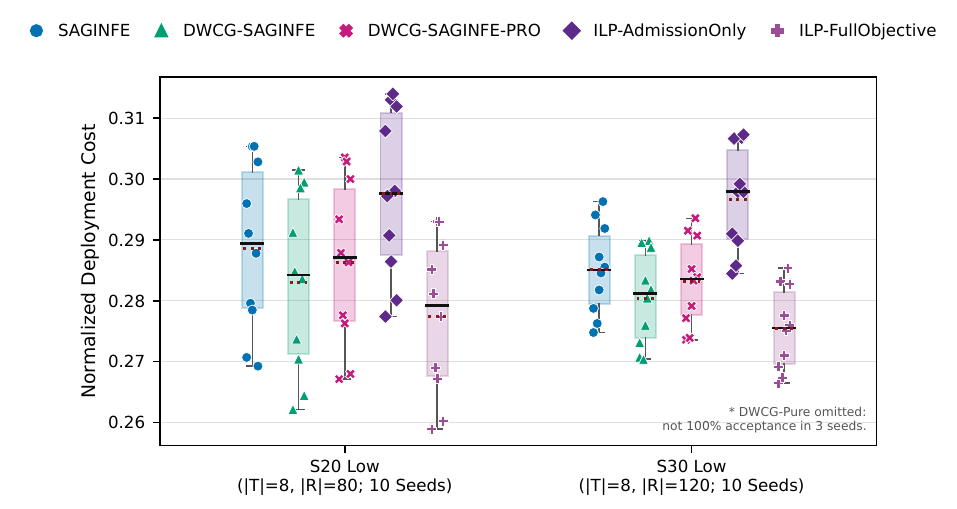}
}
\par\smallskip
{\small\textbf{(D)}~Cost: five methods, ten seeds per scale.\par}
\end{minipage}

\caption{Comparison with certified ILP references at S20 and S30. (A) Admission ratios and (B) end-to-end runtimes under Low load (left) and Medium load (right), with ten instances per available series; the runtime axes are logarithmic. ILP-AdmissionOnly certifies admission at both loads, whereas ILP-FullObjective is available only at Low load. Low-load costs are compared on (C) seven common full-admission seeds per scale for all six methods and (D) all ten seeds per scale for five methods, excluding Pure. DWCG-SAGINFE denotes HFP. Distribution summaries follow Section~\ref{subsec:simulation-setup}.}
\label{fig:results-ilp-comparison}
\end{figure*}

To examine the benefit of subsequent joint optimization directly, we compare PRO with its own SAGINFE initialization on all 120 main instances. The initial admitted count and resource cost match those of standalone SAGINFE in every case. PRO subsequently increases admission on 73 instances and retains the same admitted count with lower cost on the remaining 47. All returned results are obtained from the final integer RMP. Thus, the default configuration improves the admission-first objective on every evaluated main instance, beyond the non-worsening guarantee in Section~\ref{subsec:integer-selection}. Equal admitted counts do not imply identical admitted request sets.

\subsubsection{Comparison with Certified ILP References}

Fig.~\ref{fig:results-ilp-comparison} compares the four methods with the two ILP references on S20 and S30. At Low load, full admission makes the full-objective optimum a cost certificate for the complete request set.

Fig.~\ref{fig:results-ilp-comparison}(A) shows that the certified maximum admission ratio is 100\% throughout the tested Low- and Medium-load cases. At Low load, SAGINFE, HFP, and PRO attain this optimum on every instance, whereas Pure misses full admission on three of the ten instances at each scale. At Medium load, PRO achieves the highest mean admission among the four methods, admitting 76.9 of 80 requests at S20 and 114.9 of 120 at S30. These results leave mean gaps of 3.1 and 5.1 requests, or 3.88 and 4.25 percentage points, to the certified admission optimum. Thus, the additional optimization improves the constructive solution but does not remove the admission gap under the evaluated finite search configuration.

Fig.~\ref{fig:results-ilp-comparison}(C) includes all six methods and retains only the seven seeds per scale on which every method admits every request. On this subset, Pure's median normalized cost is 0.2853 at S20 and 0.2783 at S30, compared with 0.2812 and 0.2760 for ILP-FullObjective. The excluded instances remain part of the admission and runtime comparisons in (A) and (B).

Fig.~\ref{fig:results-ilp-comparison}(D) complements this restricted comparison by retaining all ten Low-load seeds per scale and omitting Pure. At S30, the mean normalized costs are 0.2851 for SAGINFE, 0.2832 for PRO, 0.2804 for HFP, and 0.2754 for ILP-FullObjective. PRO reduces the constructive solution's cost, but HFP achieves a lower mean cost on this full-admission set. ILP-AdmissionOnly has a higher mean cost of 0.2967, illustrating that maximum admission alone does not ensure efficient resource use. With the admitted set fixed, normalized node consumption is constant under the present model, so the deployment-dependent cost differences arise from communication-resource consumption.

Fig.~\ref{fig:results-ilp-comparison}(B) shows the computational effort required for these references. At Low load, ILP-FullObjective remains competitive, with median runtimes of 24.7 and 62.3~s at S20 and S30, compared with 49.0 and 122.1~s for PRO; SAGINFE is faster still, at 6.1 and 21.9~s. At Medium load, the median runtime of ILP-AdmissionOnly increases to 2561.1 and 11094.7~s, while PRO requires 378.2 and 488.1~s. Full-objective Medium-load runs were not pursued after one S20 diagnostic run exceeded 63~h; this unfinished run is excluded from the plotted results and provides no optimality certificate. The comparison therefore exposes a quality--time tradeoff: direct ILP optimization provides the available certificates, while finite-pool optimization obtains high admission within substantially less time in the more demanding Medium-load cases.

\subsection{Computational Scalability}\label{subsec:results-scalability}

\begin{figure}[t]
\captionsetup{font={small}}
\centering
\includegraphics[width=1\columnwidth,height=0.8\textheight,keepaspectratio]{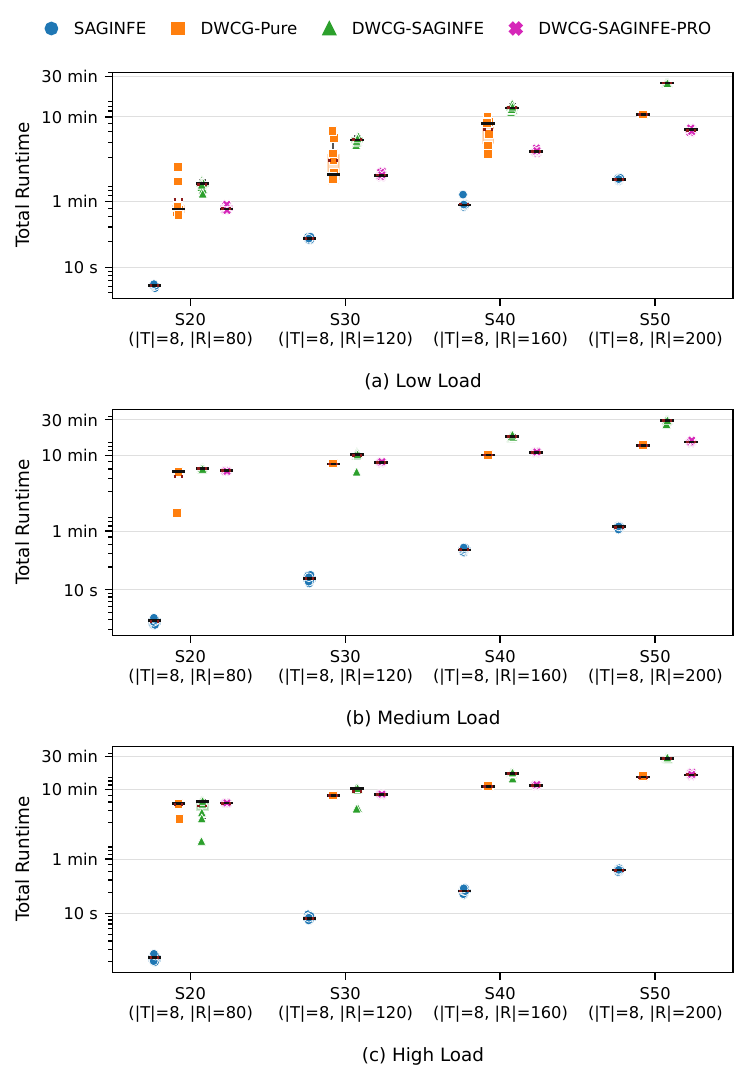}
\caption{End-to-end runtime across network scales and loads. The vertical axes are logarithmic; distribution summaries follow the conventions stated in Section~\ref{subsec:simulation-setup}.}\label{fig:results-main-runtime}
\end{figure}

Fig.~\ref{fig:results-main-runtime} reports end-to-end runtimes for the main comparison. SAGINFE consistently provides the fastest feasible solution. At S50 Medium, the median runtimes are 68.6~s for SAGINFE, 818.3~s for Pure, 906.7~s for PRO, and 1769.2~s for HFP. At S40 Low, PRO admits all requests in a median of 232.2~s, compared with 504.0~s for Pure. Initialization and the resulting column pool can thus affect both final quality and subsequent computational effort; their roles are examined further in the component study. These results support SAGINFE as a standalone deployment method when fast computation is prioritized over further improvements in solution quality. When additional computation is affordable, the configurable column-generation stage can further improve admission or resource cost, with its computational effort adjusted through pricing schedules and budgets.

\begin{figure}[t]
\captionsetup{font={small}}
\centering
\includegraphics[width=1\columnwidth,height=0.4\textheight,keepaspectratio]{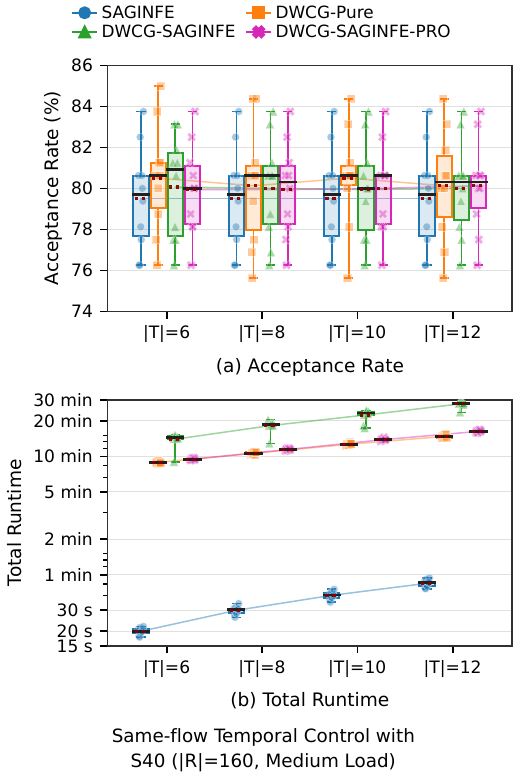}
\caption{Fixed-request temporal control at S40 Medium: admission ratio (top) and end-to-end runtime (bottom). The supplied TEG grows from 6 to 12 slots, while all request service windows remain within the first six slots.}\label{fig:results-horizon-control}
\end{figure}

The fixed-request control in Fig.~\ref{fig:results-horizon-control} increases $|T|$ from 6 to 12, doubling the number of time-expanded nodes from 240 to 480 while retaining all 160 requests and their service windows within the first six slots. Admission remains nearly unchanged: PRO's mean admitted count is 127.9 at six slots and 128.2 at twelve slots. Its median runtime increases from approximately 562 to 977~s; Pure increases from 532 to 881~s, and HFP from 868 to 1690~s. The implementation constructs models over the supplied TEG, so additional time copies increase computational work even when request windows remain fixed. This observation concerns the implemented formulation and finite search budget; normalized costs are not compared across horizons because their capacity denominators change.

\subsection{Effects of Pricing and Component Configurations}\label{subsec:results-configurations}

\subsubsection{Pricing Schedules}

We evaluate five solver-pricing schedules on S40 at Medium and High loads. EX uses strict pricing throughout, ESV starts with early-terminated pricing, and HYB-S$k$ switches from early-terminated to strict pricing after $k\in\{2,5,7\}$ rounds. ESV may switch to strict pricing if a round yields no eligible new column and sufficient budget remains, but all 40 ESV runs used only early-terminated pricing. The initialization and other components of each method remain unchanged. All evaluated runs complete 12 CG rounds and 1920 primary request-pricing calls, ending at the configured work limit. The comparison uses a common round and call budget, without asserting full master LP convergence.

\begin{table*}[t]
\small
\captionsetup{font={small}}
\caption{Component Configurations at S40: Mean Admitted Count and Median Runtime}\label{tab:results-components}
\centering
\setlength{\tabcolsep}{9pt}
\renewcommand{\arraystretch}{1.10}
\begin{tabular}{|l|r|r|r|r|r|r|}
\hline
\textbf{Configuration} & \multicolumn{2}{c|}{\textbf{Low}} & \multicolumn{2}{c|}{\textbf{Medium}} & \multicolumn{2}{c|}{\textbf{High}} \\
\cline{2-7}
 & Admitted & Time (s) & Admitted & Time (s) & Admitted & Time (s) \\
\hline
Pure & 159.2 & 504 & 148.6 & 613 & 121.8 & 670 \\
Warm & 160.0 & 230 & 151.3 & 625 & 122.4 & 643 \\
HP & 152.3 & 711 & 149.1 & 1242 & 124.4 & 1238 \\
HFP & 160.0 & 763 & 147.8 & 1068 & 121.1 & 1037 \\
PRO-no-PFE & 160.0 & 227 & 151.9 & 622 & 122.5 & 645 \\
PRO & 160.0 & 232 & 151.3 & 661 & 122.3 & 691 \\
\hline
\end{tabular}
\par\smallskip
\begin{minipage}{0.95\textwidth}
\footnotesize Each setting contains ten paired instances with 160 requests. Times include the complete algorithm and are rounded to seconds. Warm, Pure, and both PRO variants use HYB-S5; HP and HFP use heuristic-first pricing with solver fallback.
\end{minipage}
\end{table*}

\begin{figure}[t]
\captionsetup{font={small}}
\centering
\includegraphics[width=1\columnwidth,height=0.4\textheight,keepaspectratio]{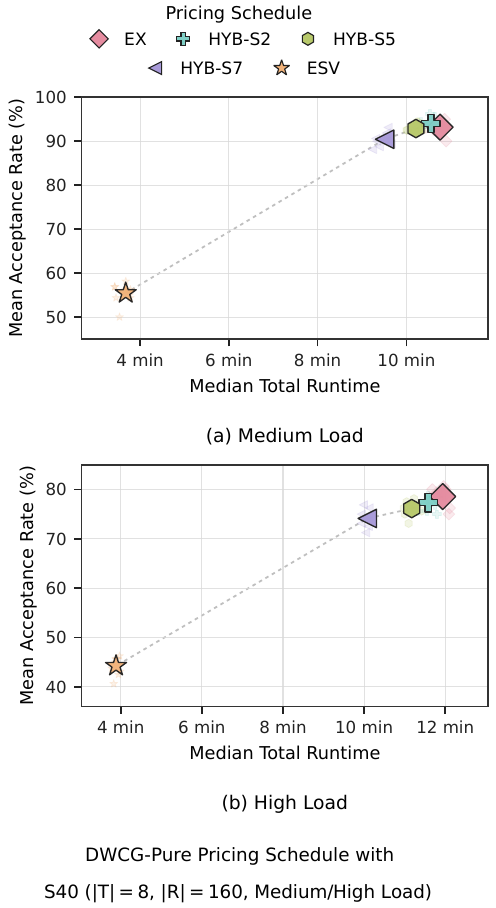}
\caption{Pricing-schedule quality--time tradeoffs for Pure at S40. The top and bottom panels correspond to Medium and High load, respectively. Large markers show mean admission ratio against median end-to-end runtime in minutes; faint points show the ten individual instances.}
\label{fig:results-pricing-schedules}
\end{figure}

\begin{figure}[t]
\captionsetup{font={small}}
\centering
\includegraphics[width=1\columnwidth,height=0.4\textheight,keepaspectratio]{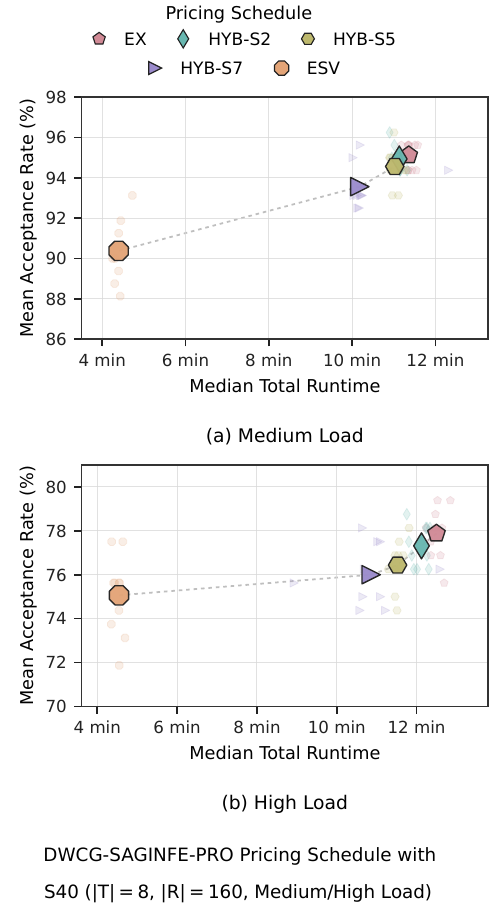}
\caption{Pricing-schedule quality--time tradeoffs for PRO at S40. The top and bottom panels correspond to Medium and High load, respectively. Large markers show mean admission ratio against median end-to-end runtime in minutes; faint points show the ten individual instances.}
\label{fig:results-pro-pricing-schedules}
\end{figure}

Fig.~\ref{fig:results-pricing-schedules} and Fig.~\ref{fig:results-pro-pricing-schedules} show the resulting quality--time tradeoffs for Pure and PRO. ESV gives the shortest runtime but fewer admissions. At Medium load, Pure-ESV admits 88.7 requests on average in 220.5~s, whereas Pure-EX admits 149.1 in 645.5~s. With SAGINFE initialization and the PRO configuration, ESV instead admits 144.6 requests in 263.7~s, and EX admits 152.2 in 681.5~s. The initial feasible deployment consequently provides a much stronger starting point when pricing is kept inexpensive. In these PRO-ESV runs, final admission equals initialization, but the returned solutions still come from integer selection over the final pool.

Hybrid schedules provide intermediate operating points. For PRO at Medium load, HYB-S7 admits 149.7 requests in a median of 611.1~s, while HYB-S2 admits 151.9 in 667.9~s. The ordering is not universally monotone: Pure-HYB-S2 admits 150.4 requests at Medium load, exceeding Pure-EX despite using slightly less time. A more thoroughly optimized pricing solution can change which alternatives enter the pool, but does not guarantee the best final integer selection under finite computation.

ESV produces smaller column pools whose final integer RMPs have median solution times below one second. In contrast, EX and the hybrid schedules typically exhaust the 300~s integer-selection allowance, with measured phase times around 302--303~s.

\subsubsection{Component Contributions}

Table~\ref{tab:results-components} examines six configurations on S40. In addition to Pure, HFP, and PRO, Warm adds SAGINFE initial columns to Pure without enabling heuristic pricing; HP denotes DWCG-Pure-HP, which starts from rejection columns and uses heuristic-first pricing with solver fallback; PRO-no-PFE disables only PRO's selective PFE calls while retaining its initialization, initial feasible solution for the integer RMP solver, and solution safeguard.

Warm initialization raises Pure's mean admission by 0.8, 2.7, and 0.6 requests at Low, Medium, and High loads, respectively. At Low load, it also reduces median runtime from 504 to 230~s. Heuristic-first pricing has a different profile: HP achieves the highest High-load mean admission in the S40 ablation, but requires approximately twice Pure's runtime and performs worse at Low load. Combining warm initialization with heuristic-first pricing does not improve every case either; HFP trails Warm at Medium and High loads. The component effects therefore depend on workload and available computation and are not uniformly additive.

Selective PFE has a modest, configuration-dependent effect. Relative to PRO-no-PFE, PRO admits 0.6 and 0.2 fewer requests on average at S40 Medium and High, while increasing runtime. The corresponding S50 comparisons instead give gains of 0.5 and 0.4 requests, again with additional runtime. At Low load, selective PFE is not invoked when initialization already admits all requests, so the small timing differences there should not be attributed to PFE calls. PFE is therefore a supplementary option whose value depends on the instance and configuration; the experiments do not identify it as the general source of PRO's improvement over initialization.

\begin{figure}[t]
\captionsetup{font={small}}
\centering
\includegraphics[width=1\columnwidth,height=0.4\textheight,keepaspectratio]{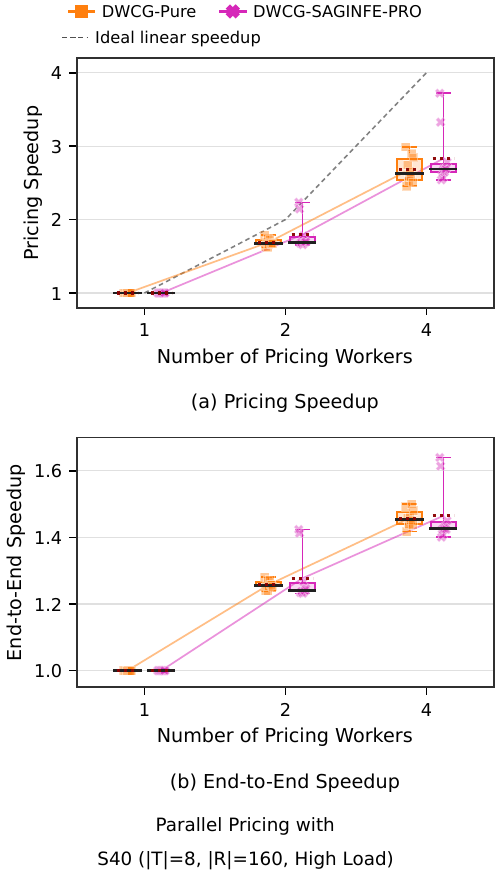}
\caption{Paired speedups with one, two, and four pricing workers at S40 High. The top panel measures primary pricing and includes an ideal linear reference; the bottom panel measures the complete algorithm. Each distribution contains ten seed-specific ratios relative to the corresponding single-worker run.}\label{fig:results-parallel-speedup}
\end{figure}

\subsection{Parallel Pricing Performance}\label{subsec:results-parallel}

Finally, we vary the number of primary-pricing workers among one, two, and four on S40 High for Pure and PRO, keeping the instance, pricing schedule, and logical work budgets fixed. Fig.~\ref{fig:results-parallel-speedup} presents the resulting speedups.

Four workers achieve median pricing-stage speedups of 2.63 for Pure and 2.69 for PRO. The corresponding end-to-end speedups are 1.45 and 1.43: median total runtime decreases from 629.0 to 433.2~s for Pure and from 662.0 to 464.5~s for PRO. Across worker counts, every paired method--instance combination retains the same admitted count, number of generated and inserted columns, and ordered final column pool.

End-to-end acceleration is smaller because initialization, master optimization, selective PFE, and final integer selection remain serial \cite{amdahl1967validity}. In particular, final integer selection stays near 302~s across worker counts and accounts for a substantial fraction of the four-worker runtime.

\section{Conclusion and Future Work}\label{sec:conclusion}

In this paper, we developed a configurable column-generation framework for full-horizon SFC deployment over predictable SAGIN topologies. An admission-first, cost-aware time-expanded model captures ordered function execution, transmission, and waiting. The framework connects SAGINFE's fast feasible construction with subsequent joint optimization, using configurable pricing and parallel execution to control computational effort. Simulations demonstrate improvements over initialization and distinct quality--time tradeoffs across configurations.

Building on learning-assisted pricing methods \cite{morabit2023arc,abouelrous2025pricing}, future work will explore RL-based single-request pricing guided by master dual prices, while retaining cross-request coordination in the RMP. Learned candidates will undergo the same feasibility and reduced-cost checks as solver-generated candidates.


%





\ifCLASSOPTIONcaptionsoff
  \newpage
\fi



\bibliographystyle{IEEEtran}
\bibliography{IEEEabrv,reference}
%

%


\makeatletter
\def\@IEEEBIOskipN{1\baselineskip}
\makeatother

\begin{IEEEbiography}
[{\includegraphics[width=1in,height=1.25in,clip,keepaspectratio]{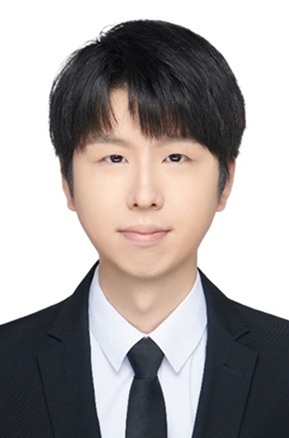}}]{Hang Li}
received the M.E. and Ph.D. degrees in information and communication engineering from Beijing University of Posts and Telecommunications (BUPT), Beijing, China, in 2019 and 2023, respectively. He is currently a Lecturer with the Department of Electronics and Communication Engineering, Yanbian University, Yanji, China. His research interests include network virtualization, network optimization, and AI-assisted resource orchestration for next-generation communication networks.
\end{IEEEbiography}

\begin{IEEEbiography}
[{\includegraphics[width=1in,height=1.25in,clip,keepaspectratio]{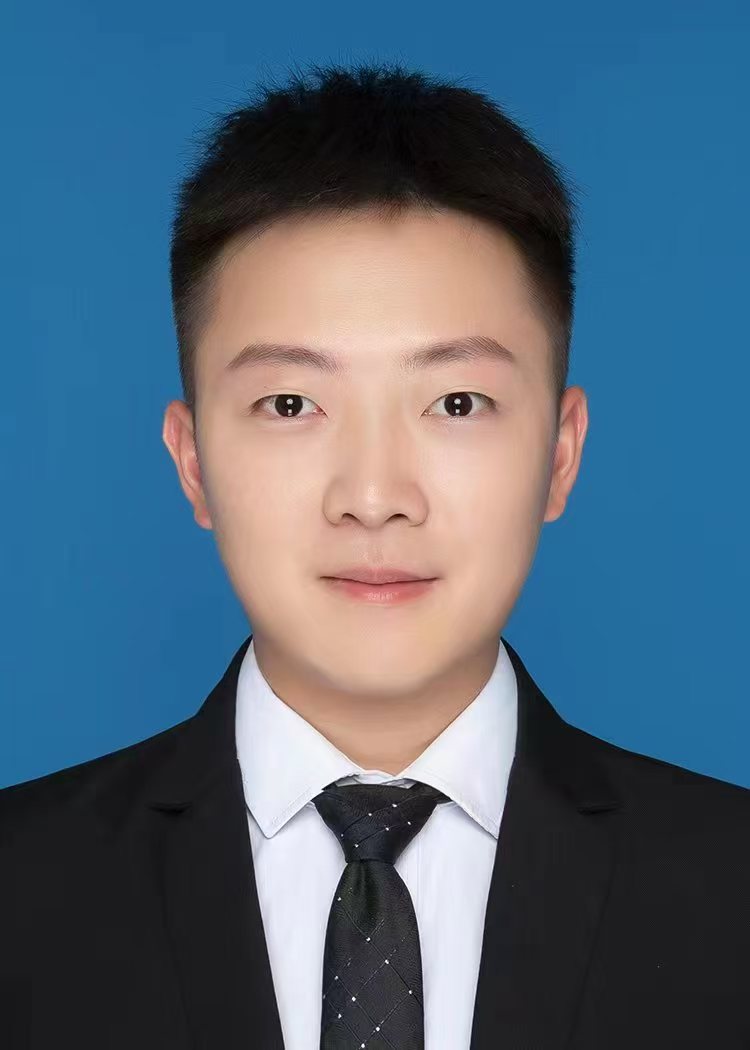}}]{Hao Huang}
received the B.E. degree in communication engineering from Chongqing University of Posts and Telecommunications (CQUPT), Chongqing, China, in 2019, and the Ph.D. degree in information and communication engineering from Beijing University of Posts and Telecommunications (BUPT), Beijing, China, in 2024. He is currently a Lecturer with CQUPT. His research interests include V2X, reinforcement learning, and intelligent transportation.
\end{IEEEbiography}

\begin{IEEEbiography}
[{\includegraphics[width=1in,height=1.25in,clip,keepaspectratio]{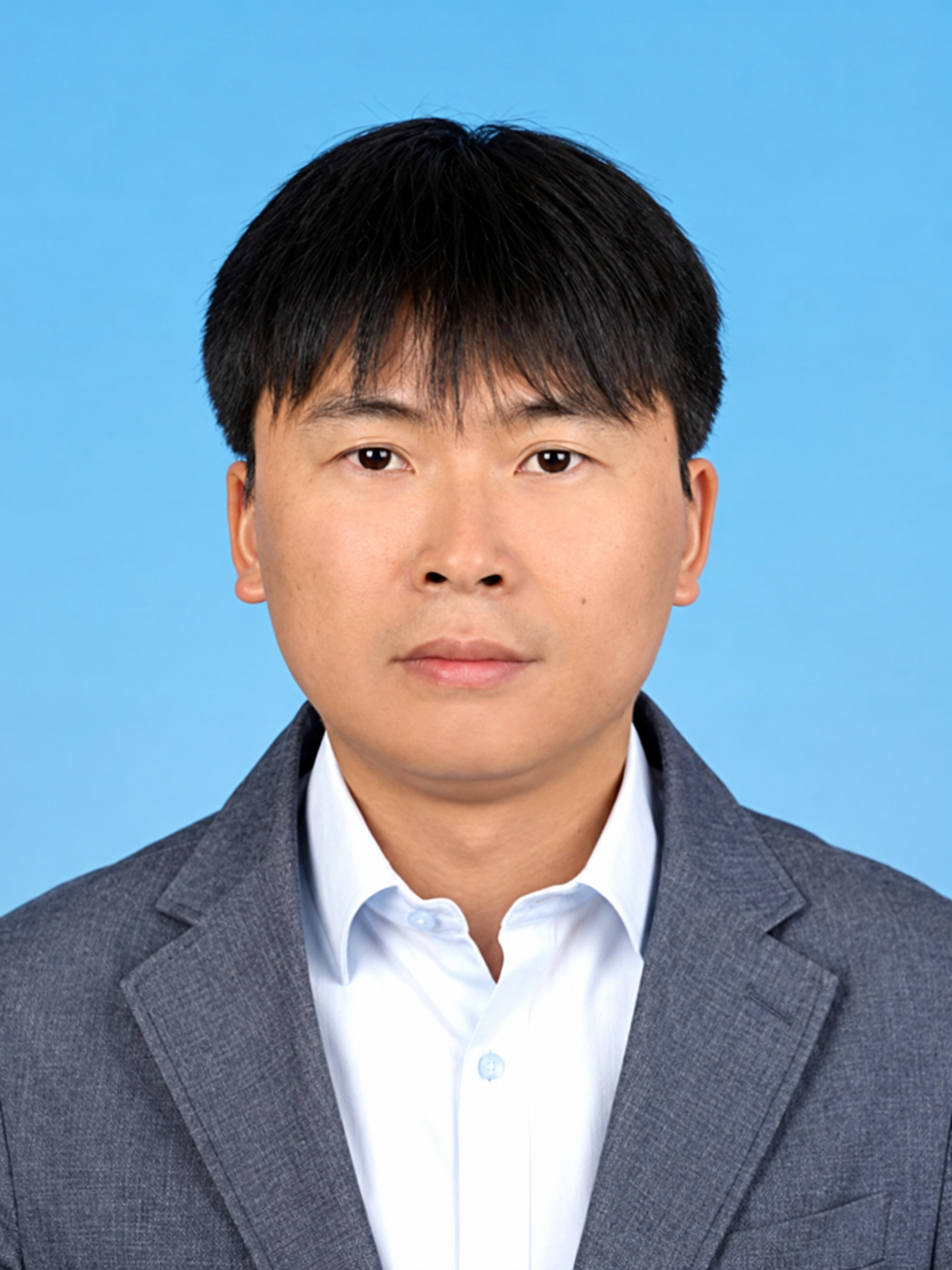}}]{Yihu Xu}
received the Ph.D. degree in electronic and information engineering from Chonbuk National University, Jeonju, South Korea, in 2015. Since 2016, he has been with the Department of Electronics and Communication Engineering,  Yanbian University, Yanji, China. His research interests include mobile communications, cognitive radio, and underwater acoustic communications.
\end{IEEEbiography}

\begin{IEEEbiography}
[{\includegraphics[width=1in,height=1.25in,clip,keepaspectratio]{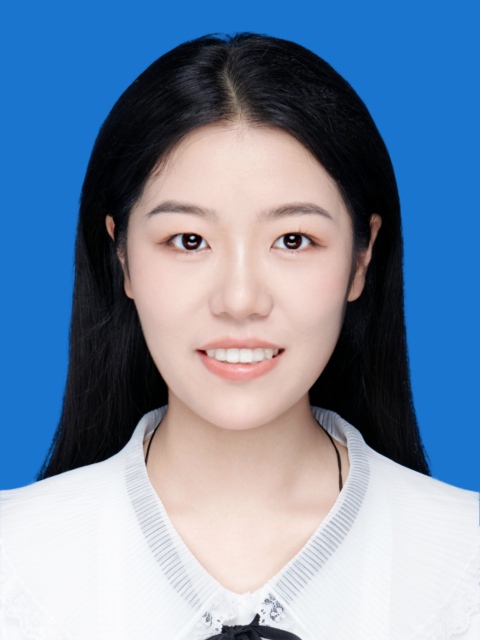}}]{Zixuan Kong}
received her M.S. degree in Information and Communication Engineering from Beijing University of Posts and Telecommunications in 2023. She is currently a Ph.D. candidate student in Information and Communications Engineering at Beijing University of Posts and Telecommunications. Her research interests include autonomous vehicles, cooperative perception, and mobile edge computing.
\end{IEEEbiography}

\begin{IEEEbiography}
[{\includegraphics[width=1in,height=1.25in,clip,keepaspectratio]{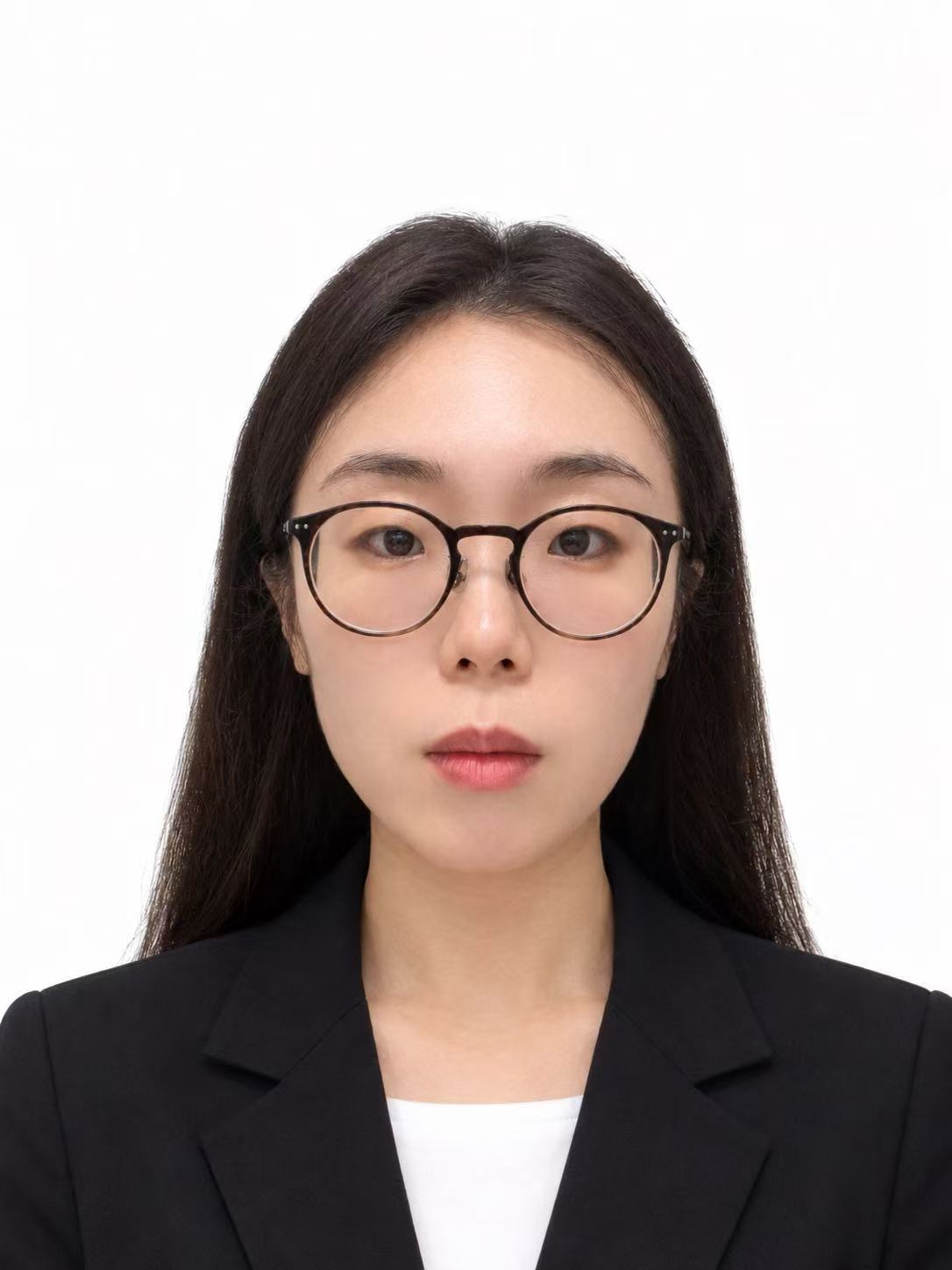}}]{Yanling Jin}
received the bachelor's degree in financial accounting from China Agricultural University, China. She received the master's degree in management accounting from Kobe University, Japan, in 2021. Since then, she has been with Fujitsu Limited, Japan, where her work focuses on data analytics and Azure-based system development. Her research interests include data analytics, demand forecasting, and cloud computing.
\end{IEEEbiography}









\end{document}